\documentclass[aps,floatfix,preprintnumbers,twocolumn,nofootinbib]{revtex4}

\usepackage{graphicx}
\usepackage{color}

\usepackage{mathrsfs}
\usepackage{hyperref}

\usepackage{amssymb}
\usepackage{amsmath}

\graphicspath{{figs/}}

\begin{document}

\title{On the spontaneous time-reversal symmetry breaking in
synchronously-pumped passive Kerr resonators}
\author{J. Rossi$^{1}$, R. Carretero-Gonz\'{a}lez$^{1,}$\footnote{{\tt URL:} http://www.rohan.sdsu.edu/$\sim$rcarrete/}, P. G. Kevrekidis$^{2}$, and M. Haragus$^{3}$.  }
\affiliation{
$^{1}$Nonlinear Dynamical Systems Group\footnote{{\tt URL:} http://nlds.sdsu.edu/}
Computational Science Research Center\footnote{{\tt URL:} http://www.csrc.sdsu.edu/}, and
Department of Mathematics and Statistics,
San Diego State University, San Diego, California 92182-7720, USA
\\
$^{2}$Department of Mathematics and Statistics, University of Massachusetts, Amherst, Massachusetts 01003-4515, USA%
\\
$^{3}$Laboratoire de Math\'{e}matiques, Universit\'{e} de Franche-Comt\'{e},  25030 Besan\c{c}on cedex, France
}

\begin{abstract}
We study the spontaneous temporal symmetry breaking instability in a coherently-driven passive optical Kerr resonator observed experimentally by Xu and Coen in Opt.~Lett.~{\bf 39}, 3492 (2014). We perform a detailed stability analysis of the Lugiato-Lefever model for the optical Kerr resonators and analyze the temporal bifurcation structure of stationary symmetric and the emerging asymmetric states as a function of the pump power. For intermediate pump powers a pitchfork loop is responsible for the destabilization of symmetric states towards stationary asymmetric ones while at large pump powers we find the emergence of periodic asymmetric solutions via a Hopf bifurcation. From a theoretical perspective, we use local bifurcation theory in order to  analyze the most unstable eigenmode of the system. We also explore a non-conservative variational approximation  capturing, among others, the evolution of the solution's amplitude, width and center of mass. Both methods provide insight towards the pitchfork bifurcations associated with the symmetry breaking.
\end{abstract}


\maketitle

\section{Introduction}
Spontaneous symmetry breaking (SSB) is the basis for many phase transitions and account for effects including ferromagnetism, superconductivity, and convection cells~\cite{zurek,stanley}.
SSB has been widely observed in nonlinear optics and is at the heart of numerous
fundamental phenomena including, but not limited to, asymmetric
dynamics in coupled 
mode models~\cite{kenkre}, optical waveguide arrays~\cite{Boris_Gisin_Kaplan_PRE_2000},
coupled nonlinear micro-cavities~\cite{Haelterman_OE_2006}, 
and photonic lattices~\cite{Panos_PLA_2005}. For a detailed exposition
of numerous recent directions within the subject from the perspective
of nonlinear phenomena, see Ref.~\cite{Boris_SSB_book}.
SSB is not restricted to Hamiltonian (conservative) systems. For instance, 
over the past few years,
it 
has also played a prominent role in the context of parity-time, 
so-called {\cal PT}, symmetric
systems~\cite{kivshar,yang} bearing a balanced interplay between 
gain and loss. 
There, it is responsible for the emergence of
``ghost'' states both in the case of dimers
(and more generally oligomers)~\cite{cartarius}, but also in that
of continuous media~\cite{rcg:85,rcg:88}, where they can be responsible
for the destabilization and bifurcations associated with solitary
waves and vortices.

A remarkable example of SSB in a dissipative system was observed by 
Xu and Coen in Ref.~\cite{XuCoen} where a system composed of 
a synchronously-pumped 
passive optical resonator filled with a Kerr nonlinear material was experimentally 
explored. This system exhibits a temporal SSB instability in which the 
discrete time-reversal symmetry is broken and symmetric states become unstable
in favor of stable asymmetric states.
It is the purpose of the present manuscript to complement
the experimental and numerical analysis of Ref.~\cite{XuCoen} by
putting forward a thorough analytical
(and partially numerically assisted) 
understanding of the origin and manifestation of SSB 
(and additional possible, such as Hopf) bifurcations in this 
system.

We consider, as in Ref.~\cite{XuCoen}, a model for a passive Kerr resonator in an 
optical fiber ring cavity described by a single partial differential equation (PDE),
resulting from an averaging procedure, of the nonlinear Schr\"odinger (NLS) 
equation-type, known as the mean-field Lugiato-Lefever (LL) 
model~\cite{LL,LLE}. The LL
equation, taking into account gain and loss in the system, can be cast, 
in non-dimensional form, as~\cite{XuCoen,XuCoenRef22a,XuCoenRef22b}:
\begin{equation}
\frac{\partial E(z, \tau)}{\partial z} = \left[ -1 + i (\left\vert E\right\vert^2 - \Delta ) - i \eta \frac{\partial^2}{\partial \tau^2} \right] E + S(\tau), 
\label{NLSEq}
\end{equation}
where $z$ is the slow evolution variable of the intracavity field $E$ over 
successive normalized cavity round-trips and $\tau$ describes the temporal 
variable in the dependence of the intracavity pulse envelope.  
The terms in the right-hand-side  of Eq.~(\ref{NLSEq}) correspond, respectively, to
cavity losses $(-E )$, Kerr nonlinearity $(i \left\vert E\right\vert^2 E)$, cavity 
phase detuning $(-i \Delta E)$, chromatic dispersion 
$( - i \eta \frac{\partial^2}{\partial \tau^2}E)$, and external pumping $( S(\tau))$.  
Within this non-dimensional form~\cite{XuCoenRef22a,XuCoenRef22b}, 
the cavity phase detuning
corresponds to $\Delta = \delta_0\alpha$, where $\alpha$ is half the fraction of 
power lost per round-trip and the cavity finesse is $\mathscr{F} = \pi/\alpha$, and 
$\delta_0 = 2m\pi  - \phi_0$ where $\phi_0$ is the overall cavity round-trip 
phase shift and $m$ is the order of the closest cavity resonance.  
The sign of the group-velocity dispersion coefficient of the fiber is $\eta$ 
which is taken as $\eta = -1$ for our analysis with self-focusing nonlinearity.  
The field envelope of the external pump pulses, $S(\tau)$, is modeled by a symmetric chirp-free Gaussian pulse given by $S(\tau) = \sqrt{X} \exp\left[ -(\tau/T_0)^2 \right]$,
with $T_0 = 2.3$ as in the experiments of Ref.~\cite{XuCoen}.

For the SSB instability of the passive Kerr cavity, the pump pulse field profile is temporally symmetric, $S(\tau) = S(-\tau)$, and the model is symmetric under a time reversal transformation, $\tau \rightarrow -\tau$, 
yet it admits asymmetric solutions, as described in Ref.~\cite{XuCoen}.  The 
associated pitchfork bifurcation illustrates that at low pump peak power $X$, the solutions are symmetric in time; however, above a certain pump peak power threshold the symmetric states become unstable while stable asymmetric states
emerge.  
The particular experimental parameters of Ref.~\cite{XuCoen} generate, as $X$ is
increased further, a reverse pitchfork as well, in which the asymmetric states
collide and disappear while the symmetric state recovers its stability.
We examine this SSB-induced 
instability interval in the passive 
Kerr resonator modeled by Eq.~(\ref{NLSEq}) by means of a non-conservative variational
approximation (NCVA)~\cite{JuliaNCVA} and further through
a center manifold reduction~\cite{MH-Iooss}
enabling the analysis of the dominant associated eigenmodes (responsible
for determining the spectral stability of the system).  
It is relevant to mention at this point that a thorough  bifurcation analysis for a LL equation in the case of constant external pumping
was recently carried out in Ref.~\cite{LLE_french_PRA}, showing quite complex bifurcation scenarios in both the anomalous and normal dispersion regimes.

In the NCVA context, our aim is to
apply a variational method based on well-informed ans\"{a}tze 
in the corresponding
Lagrangian of the system.  The ans\"{a}tze reduce the complexity of the 
original 
infinite-dimensional problem to a few degrees of freedom capturing the 
principal,
static and dynamic characteristics of the system.  
This method attempts  to project the infinite-dimensional dynamics of
Eq.~(\ref{NLSEq}) into a low-dimensional dynamical system that qualitatively and, to some
extent, quantitatively captures SSB bifurcations and the solutions emanating from it.
However, it is important to note that traditional variational methods rely on the 
existence of a Lagrangian or Hamiltonian structure for closed systems for which 
equations of motion can be derived. Nonetheless, recently, Galley~\cite{ref1} offered 
an approach allowing to extend the method to open, 
non-conservative  systems 
which in turn was generalized to dissipative (containing gain and loss) NLS-type
systems in Ref.~\cite{JuliaNCVA} inspired by the work of Ref.~\cite{ref4} on the 
extension of Galley's formalism to {\cal PT}-symmetric variants of field theories. It is this variant of the NCVA that we will explore in the present
setting.

{Our analysis of the observed SSB will be complemented by center 
manifold reductions.
The latter are extensively used in the analysis of local bifurcations. Starting from a dynamical systems formulation of the bifurcation problem, the reduction to a center manifold provides the lowest dimensional dynamical system which fully describes the original dynamics close to a bifurcation point. We use this method to analyze the two pitchfork bifurcations which arise in  Eq.~(\ref{NLSEq}) as $X$ is increased. As a result we obtain, in both cases, a reduced scalar ordinary differential equation which captures the bifurcating dynamics. The first two coefficients in the expansion of the reduced scalar field, which are computed numerically here, determine the type of the bifurcation. They are also essential in the computation of the bifurcating asymmetric states and of the local temporal dynamics. 
}

The paper is organized as follows.
In Sec.~\ref{secModel} we identify the equilibria and study their stability by means 
of a spectral analysis of the linearization problem; this is a perspective
that was absent in the original work of Ref.~\cite{XuCoen} and which, we 
argue, provides a more systematic insight into the stability (and the
potential instabilities) of the system. 
In doing so, we recover the forward and reverse 
pitchfork bifurcations (i.e., a pitchfork loop) observed in Ref.~\cite{XuCoen} as well 
as identify a Hopf bifurcation for larger pump power giving rise to asymmetric,
stable, periodic solutions; the latter is an important feature of 
dynamical interest
in its own right and should be, in principle, observable in suitable
extensions
of the experiments of~\cite{XuCoen}.
Section~\ref{secNCVA} is devoted to the NCVA approach.
In Sec.~\ref{secNCVA:prelim} we provide a brief description of the 
NCVA approach and its formulation within the LL model.
Section~\ref{secNCVA:app} is devoted to the application of the NCVA to capture
the SSB bifurcation for physically relevant parameters values of the system 
as in Ref.~\cite{XuCoen}.
In Sec.~\ref{secGalerkin} we complement our understanding of the pitchfork
loop bifurcation by giving {the local bifurcation analysis} which is effective 
towards qualitatively and quantitatively describing the emerging asymmetric solutions
close to the pitchfork bifurcation points. 
Finally, in Sec.~\ref{secConclusion} we summarize
 our findings and we provide possible
avenues for future research.

\section{The Full Model: Equilibria, Stability and Bifurcations
\label{secModel}}
In this section, we follow the various equilibria of Eq.~(\ref{LugiatoLefever}) as the 
peak pump power, $X$, is varied and determine their stability.  
Let us recast Eq.~(\ref{NLSEq}) into the simpler form
\begin{equation}
 i u_z + u_{\tau \tau} + (|u|^2 - \Delta) u = - i u  + i S(\tau),
 \label{LugiatoLefever}
\end{equation}
which corresponds to the NLS with additional non-conservative terms
(namely the terms in the right-hand side).
In what follows, we identify stationary solutions, $u(z,\tau)=u_0(\tau)$
of Eq.~(\ref{LugiatoLefever}) by numerically solving the steady-state equation
\begin{equation}
 u_{0,\tau \tau} + (|u_0|^2 - \Delta) u_0 = - iu_0  + i S(\tau).
\label{u0}
\end{equation}
It is relevant to mention that since the forcing 
(pump) term in Eq.~(\ref{NLSEq}) is independent
of the field's wavefunction, it is necessary for the steady state to 
be independent of $z$ (i.e., here the detuning parameter $\Delta$ 
plays the role of the frequency). It is also 
worth mentioning that the steady state
is, in general, complex which, as we will see below, is crucial for the steady
state to sustain itself through a stationary flow from the gain to the loss
portions of the solution.  

Let us now consider the {stability of a steady state} $u_0$ by means 
of a spectral stability analysis.
Specifically, small perturbations of order $\mathcal{O}(\epsilon)$, 
with $0<\epsilon \ll 1$, to the stationary solutions are introduced in the 
form:
\begin{equation}
u(z, \tau) =  u_0(\tau) + \epsilon [ a(\tau)e^{\lambda z} + b^*(\tau)e^{\lambda^* z}],
\nonumber
\end{equation}
and substituted into Eq.~(\ref{LugiatoLefever}).  Then, the ensuing linearized equations are solved to $\mathcal{O} (\epsilon)$, leading to the eigenvalue problem:
\begin{align}
i \lambda \begin{pmatrix} a(z) \\ b(z) \end{pmatrix} = \begin{pmatrix} M_1 & M_2 \\  -M_2^* & -M_1^* \end{pmatrix} \begin{pmatrix} a(z) \\ b(z) \end{pmatrix}, 
\label{PDEEigenProblem}
\end{align}
for the eigenvalues $\lambda$ and associated eigenvector $\xi = (a(z),b(z))^\mathrm{T}$,
where $(\cdot)^*$ denotes complex conjugation and $M_1$ and $M_2$ are the following
operators:
\begin{align}
M_1 &=  - \partial_\tau^2 - 2|u_0|^2+(\Delta - i), \nonumber \\
M_2 &= -u_0^2.
\end{align}
The stationary solutions are linearly unstable provided Re$(\lambda)> 0$.
When unstable, the dynamics of the respective instabilities can be monitored 
through direct numerical simulations of Eq.~(\ref{LugiatoLefever}).
%

\begin{figure}[htb!]
\centering
\includegraphics[width=8.5cm]{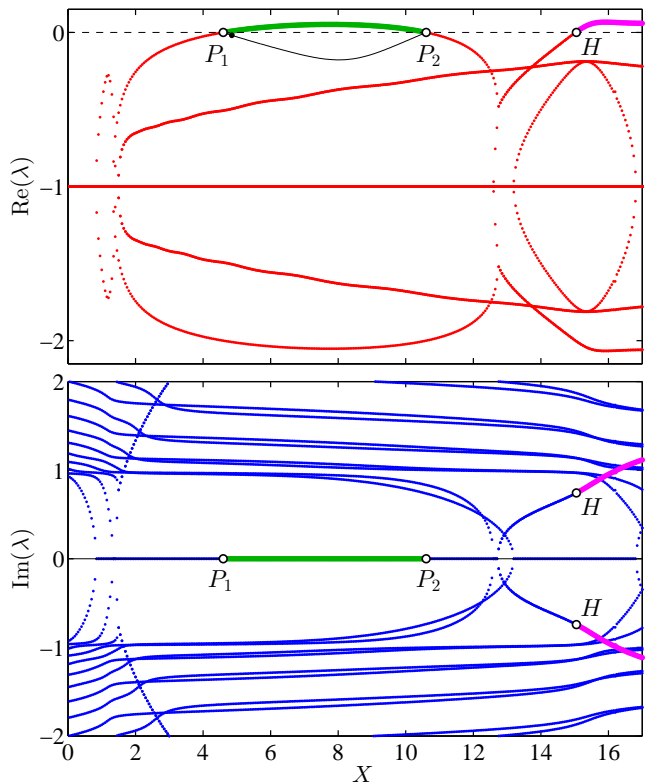}
\caption{Linearization spectrum for the symmetric and asymmetric 
steady state solutions of 
the Lugiato-Lefever equation (\ref{LugiatoLefever})
as the pump power $X$ is varied for $\Delta = 0.92$ and $T_0=2.3$.  
The top and bottom panels depict, respectively, the real and imaginary
parts of the eigenvalues.
Stable symmetric solutions bearing Re$(\lambda)<0$ are depicted by 
small (red) dots in the top panel while unstable symmetric solutions 
are depicted with thick solid lines.
The thick (green) solid line between the points $P_1$ and $P_2$ represents
the unstable solutions through forward ($P_1$) and reverse ($P_2$)
pitchfork bifurcations.
The thin (black) curve between the points $P_1$ and $P_2$ corresponds to the 
stable asymmetric solution branches created through the pitchfork bifurcation.
(The small black dot next to the point $P_1$ is the stable eigenvalue
used for the slope computation in Fig.~\ref{PitchPhasePortrait}.)
The thick (magenta) solid line to the right of the Hopf bifurcation point
$H$ indicates the onset of instability for the symmetric state and the 
existence of an asymmetric periodic solution.
}
\label{fig:frequencySpectrum}
\end{figure}

\begin{figure}[t]
\centering
\includegraphics[width=8cm]{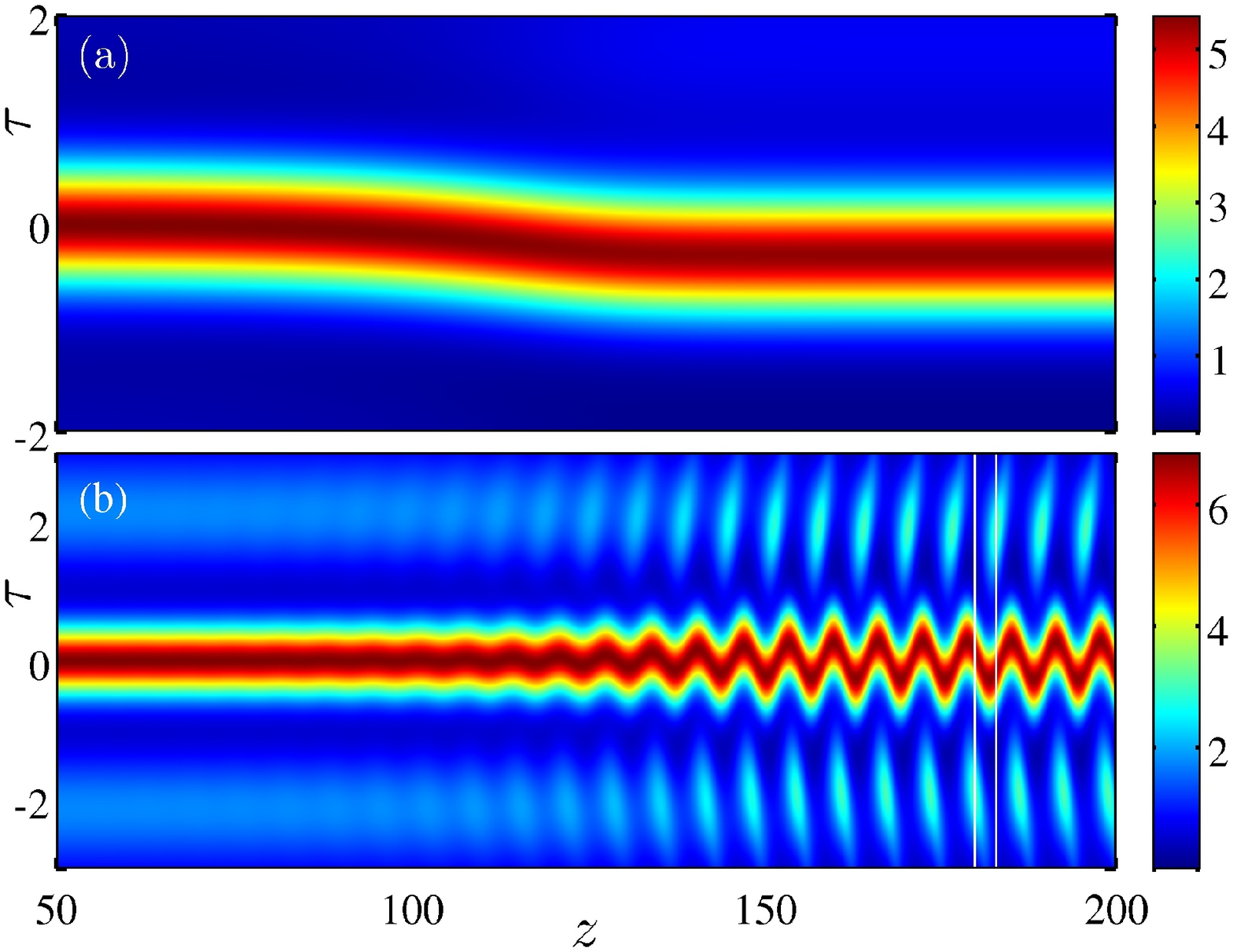}\\
\includegraphics[width=8cm]{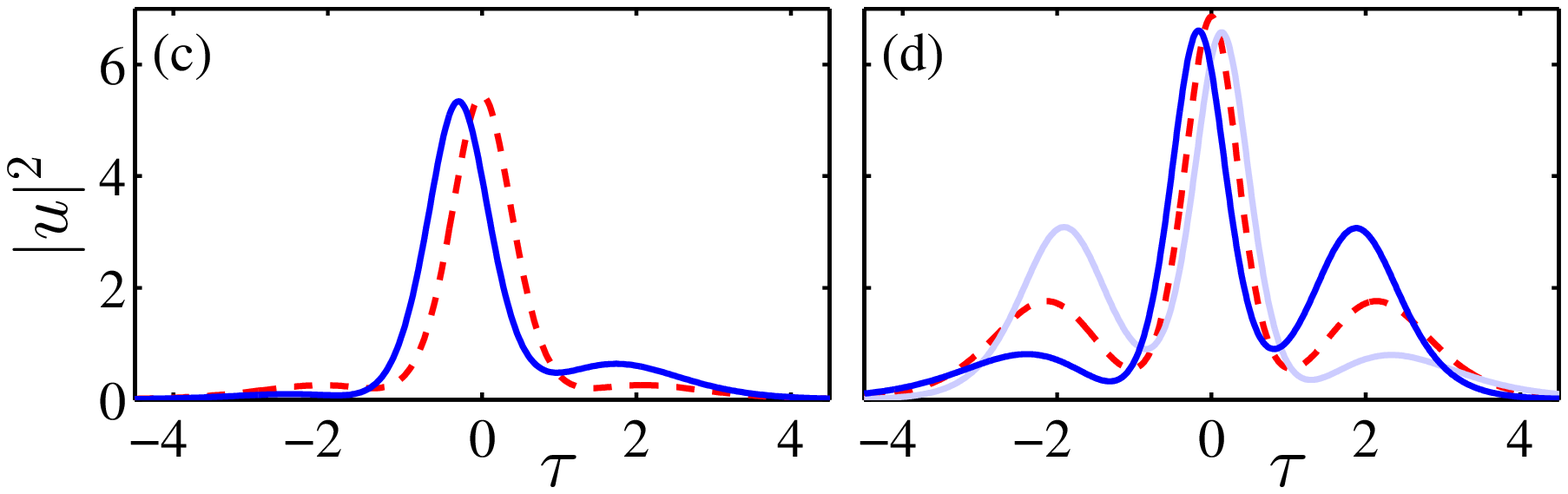}
\vspace{-0.3cm}
\caption{
(a), (b) Examples for the density evolution of unstable symmetric states
and 
(c), (d) snapshots for the corresponding states.
(a) Evolution of the unstable symmetric state for $X=8$ between the two pitchfork bifurcations 
$P_1$ and $P_2$ depicted in Fig.~\ref{fig:frequencySpectrum}. The initial symmetric state,
see dashed (red) line in panel (c) evolves towards the asymmetric steady state
depicted in solid (blue) in panel (c).
(b) Evolution of the 
unstable symmetric state towards a periodic breathing solution
for $X=16$ (i.e., to the right of the Hopf bifurcation point $H$ in 
Fig.~\ref{fig:frequencySpectrum}). The initial symmetric state [dashed (red)
line] and two snapshots of the density for the periodic solution [solid (blue and 
light blue) lines] separated by half a period, at the times corresponding to the white
vertical lines in panel (b), are depicted in panel (d).
\label{fig:evolution}}
\end{figure}

Figure~\ref{fig:frequencySpectrum} depicts the linearization 
spectrum for the symmetric stationary
solution [see (red) dashed line in panels (c) and (d) of Fig.~\ref{fig:evolution}] 
as a function of the pump peak power.  
The spectrum in Fig.~\ref{fig:frequencySpectrum} evidences the existence of two 
unstable branches:
(i) a pitchfork bifurcation loop containing a forward pitchfork 
bifurcation, see point $P_1$ at $X\approx 4.6$, and a reverse 
pitchfork bifurcation, see point $P_2$ at $X\approx 10.6$, and 
(ii) a Hopf bifurcation, see point $H$ at $X\approx 15.1$.
The pitchfork bifurcation, see thick (green) line between the 
points $P_1$ and $P_2$ in Fig.~\ref{fig:frequencySpectrum}, 
is responsible, as the pump power is increased, for the loss of stability of the 
symmetric state towards a pair of asymmetric states (one to the left and one to the 
right) at $P_1$. As the pump power is increased, a reverse pitchfork at $P_2$
is responsible for the collision (and annihilation) of the two asymmetric states 
towards the symmetric state that recovers its stability.
A sample of the dynamic destabilization of the (unstable) symmetric state
for a pump strength $X=8$, namely between the two pitchfork points, is depicted 
in Fig.~\ref{fig:evolution}(a). As the figure shows, the symmetric state 
[see dashed (red) line in Fig.~\ref{fig:evolution}(c)] destabilizes
towards the stable, asymmetric state [see solid (blue) line in
Fig.~\ref{fig:evolution}(c)].
On the other hand, the instability due to the Hopf bifurcation branch,
see the thick (magenta) line emanating from the point $H$ in
Fig.~\ref{fig:frequencySpectrum}, is responsible for the instability of
the symmetric state towards a {\em periodic} (in $z$) solution. A sample of the
evolution for the symmetric state towards the stable periodic solution
is depicted in Fig.~\ref{fig:evolution}(b). The periodic solution contains
three ``humps'' in its $\tau$ dependence: 
a central one performing left-to-right oscillations while
the side ``humps'' oscillate alternatively up-and-down. Snapshots for the
asymmetric states when the side ``humps'' have the largest magnitude are
depicted in panel (d) corresponding to the times depicted by a horizontal
white line in panel (b).

It is important to mention that, due to the cavity loss term ($-iu$), the
real part of the spectrum is symmetric with respect to Re$(\lambda)=-1$
(see Sec.~\ref{secGalerkin} for details).
Therefore, tuning the cavity loss parameter is crucial to the existence of the
SSB bifurcation as higher values of this parameter shift the real part of the
spectrum down precluding the possibility of eigenvalues crossing the
origin and leading to such bifurcations. By the same token reducing the value
of the cavity loss parameter will induce more eigenvalues to cross the
origin and thus leading to richer and more complicated bifurcation scenarios.
A detailed analysis of the bifurcations as the cavity loss parameter is
varied is outside of the scope of the present manuscript and will be
studied in a future work.

\section{Non-conservative Variational Approximation
\label{secNCVA}}
\subsection{Preliminaries
\label{secNCVA:prelim}}

To employ the NCVA, we consider two sets of dependent
variables $u_1$ and $u_2$.  As proposed by Galley and collaborators~\cite{ref1,Galley:14}, these are fixed at an initial time ($z_i$), but are not fixed at the final time ($z_f$).  After applying variational calculus for a non-conservative system, both paths are set equal, $u_1= u_2$, and identified with the physical path $u$, the so-called physical limit (PL).  The action functional for $u_1$ and $u_2$ is defined as the total action integral of the difference of the Lagrangians between the  paths plus the action integral of the functional ${\cal R}$ which describes the generalized non-conservative forces and depends on both paths:
\begin{eqnarray}
S =&  \int_{z_i}^{z_f}  & dz \int_{-\infty}^{\infty} d\tau [\mathcal{L} (u_1, u_{1,z}, u_{1,\tau}, \ldots,z)  \\ \nonumber
&& - \mathcal{L}(u_2, u_{2,z}, u_{2,\tau}, \ldots,z) + \mathcal{R} ],
\end{eqnarray}
where the $z$ and $\tau$ subscripts denote partial derivatives
with respect to these variables.
The above action defines a new total Lagrangian density:
\begin{equation}
\mathcal{L}_T \equiv  \mathcal{L}_1 - \mathcal{L}_2  + \mathcal{R},
\label{eq:action}
\end{equation}
where the first two terms represent the conservative Lagrangian densities for which $\mathcal{L}_i \equiv \mathcal{L}(u_i, u_{i,z}, u_{i,\tau}, ...,z)$, for $i=1,2$, and $\mathcal{R}$ contains all the non-conservative terms.
For convenience, $u_+ = (u_1 + u_2)/2$ and $u_- = u_1 - u_2$ are defined in such a way that at the physical limit $u_+ \,  \rightarrow \, u$ and $u_- \, \rightarrow \, 0$.  
%
Then, the modified Euler-Lagrange equations for the effective Lagrangian $L = \int_{-\infty}^\infty \mathcal{L}_T d\tau$ yield
\begin{equation}
\frac{\partial L}{\partial u} - \frac{d}{dt}\left( \frac{\partial L}{\partial \dot{u}} \right) + \int_{-\infty}^\infty \left[ \frac{\partial \mathcal{R}}{\partial u_- }\right]_{\mathrm{PL}} d\tau = 0. 
\label{NCVAODE}
\end{equation}
Through this method we recover the Euler-Lagrange equation for the conservative terms and all the non-conservative terms are folded into $[ \frac{\partial \mathcal{R}}{\partial u_-} ]_{\mathrm{PL}}$.  
It is crucial to construct the term $\mathcal{R}$ such that its derivative with respect to the difference variable $u_-=u_1-u_2$ at the physical limit gives back the non-conservative or generalized forces. This part concludes
the field-theoretic formulation of the non-conservative problem and
so far no approximation has been utilized. The latter
will stem from the use of an approximate ansatz for
the solutions within the variational method
for this extended (to the non-conservative case) Lagrangian formulation.
One key aspect of any variational method is the proper, judicious, choice
of ansatz.  In this paper, we compare two different ans\"{a}tze with 
four and six parameters (i.e., degrees of freedom).  
We apply the NCVA to Eq.~(\ref{LugiatoLefever}) to verify if the reduced
dynamical system is able to qualitatively (and quantitatively) capture the 
SSB instability by following all temporally symmetric and asymmetric solutions 
to the reduced system of ODEs given by Eq.~(\ref{NCVAODE}).
The conservative Lagrangian density for the NLS, namely Eq.~(\ref{LugiatoLefever})
with the right-hand-side equal to zero, is
\begin{equation}
\mathcal{L} = \frac{i}{2} \left(u^* u_z- u u_z^*\right) -  \left\vert u_{\tau}\right\vert^2 +  \frac{1}{2}\left\vert u \right\vert^4 - \Delta \left\vert u \right\vert^2.
\end{equation}
Here, we construct $\left[ \partial \mathcal{R}/\partial u_- \right]_{\mathrm{PL}} = - iu + i S(\tau)$, by choosing $\mathcal{R} = \left( -iu_{+} + i S(\tau) \right)u_-$.  Therefore, the relevant non-conservative Lagrangian density can be written as
\begin{eqnarray}
\mathcal{L} &=& \frac{i}{2} \left(u_{1}^* u_{1,z}- u_{1} u_{1,z}^*\right) -  \left\vert u_{1,\tau}\right\vert^2 + \frac{1}{2} \left\vert u_1 \right\vert^4 -  \Delta  \left\vert u_1 \right\vert^2  \nonumber
 \\
&-& \frac{i}{2} \left(u_{2}^* u_{2,z}- u_{2} u_{2,z}^*\right) +  \left\vert u_{2,\tau}\right\vert^2 -\frac{1}{2}  \left\vert u_2 \right\vert^4 +  \Delta  \left\vert u_2 \right\vert^2  \nonumber \\ 
&+& \left( - iu_{+} + i S(\tau) \right)u_-,  
\end{eqnarray}
where $u_1 = \left( 2u_+ + u_- \right)/2$ and $u_2 = \left(2u_+ - u_- \right)/2$.
For reasons of brevity, we chose to express the Lagrangian density above
in 1,2 coordinates. Writing the Lagrangian in $\pm$ coordinates lends
itself to lengthier expressions but to more straightforward implementation
of the physical limit where the (+) variables directly coincide with the
physical variables [and the $(-)$ variables are eliminated].
From $\mathcal{L} $, we can derive, through the Euler-Lagrange
equations (\ref{NCVAODE}), the full LL model at the PDE level.
In order to, however, obtain an analytical insight in the dynamics
of the model, our aim is to use an ansatz approximation of the
pulse reducing its Lagrangian to a Lagrangian over effective
(yet time-dependent) properties of its form, like the amplitude,
the width and its center of mass, among others. Then for these
effective properties, in the spirit of Refs.~\cite{JuliaNCVA,ref4},
a coupled system of 
ODEs approximating
the dynamical evolution will be derived and, perhaps more
importantly for our considerations, their corresponding steady
states and possible bifurcations will be amenable to 
analysis.

\subsection{Bifurcation Analysis Using the NCVA Approach
\label{secNCVA:app}}
%
\begin{figure*}[htb!]
\centering
\includegraphics[width=15cm]{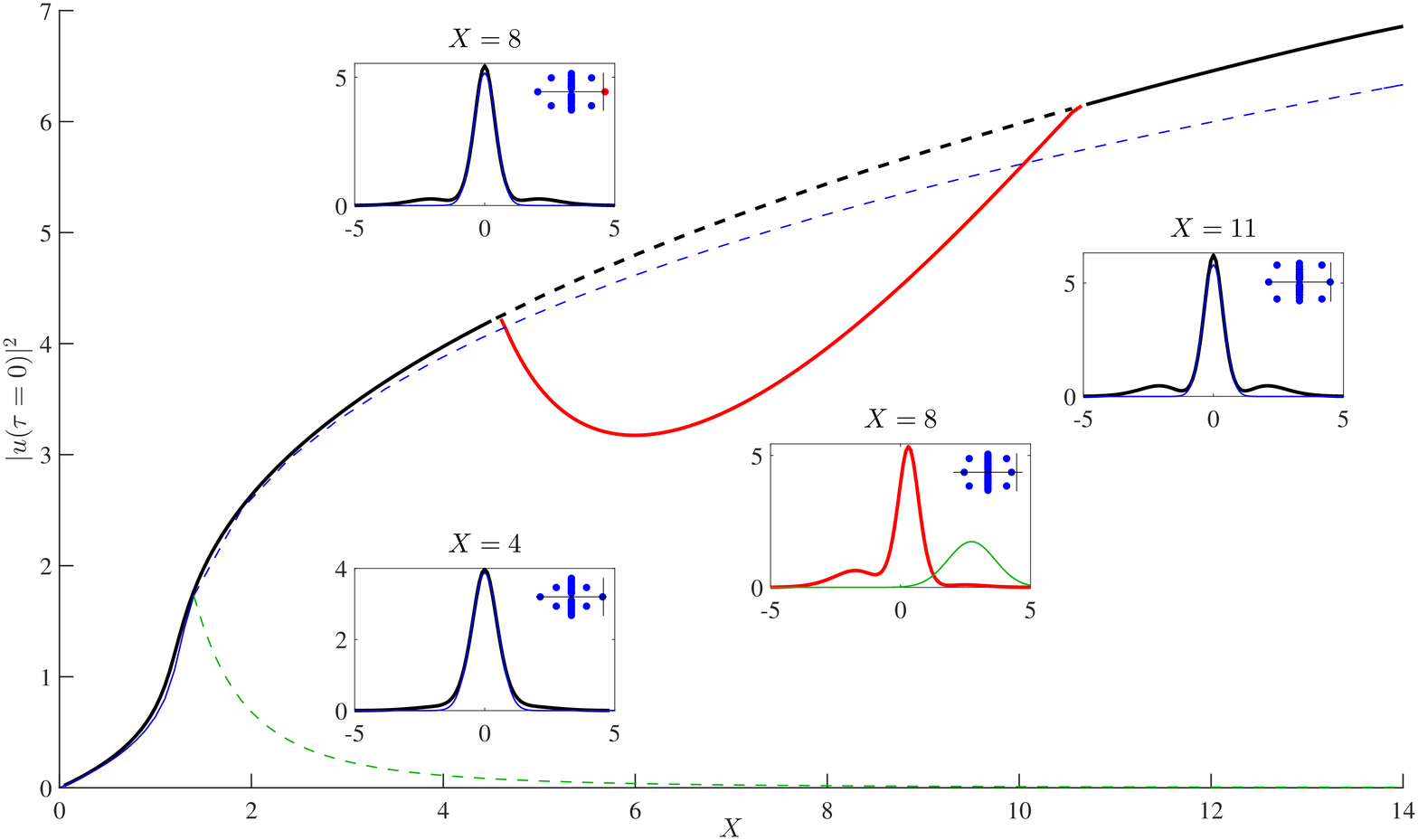}
\caption{Bifurcation for steady states of the LL model (\ref{LugiatoLefever})
(thick curves) and their approximation using the NCVA methodology with the 
over-simplified four-parameter ansatz (\ref{eq:4pAnsatz}) (thin curves)
as the pump strength $X$ is varied for $\Delta = 0.92$ and $T_0 = 2.3$.
Stable (Unstable) branches are depicted with solid (dashed) lines.
The (red and green) branches bifurcating from the main branch (blue and 
black lines) correspond to asymmetric solutions.
The insets depict the pulse temporal intensity profiles obtained for 
$X=4$ (symmetric), $X=8$ (symmetric and asymmetric), and $X=11$ (symmetric) for
the original LL model (thick curves) and their NCVA approximation
(thin curves).
The insets also depict the corresponding stability spectra
for these solutions where stable eigenvalues are depicted in blue 
and unstable eigenvalues in red.
\label{figSSB4p}}
\end{figure*}

Applying the NCVA methodology described above to the LL model
(\ref{LugiatoLefever}) where $\left[ \partial \mathcal{R}/\partial
u_- \right]_{\mathrm{PL}} = -iu  + i S(\tau)$, yields 
$\mathcal{R} = \left( -iu_{+}  + i S(\tau) \right)u_-$.  
We first choose the following simple Gaussian ansatz
\begin{equation}
\bar{u}_j = a_j \exp\left[-\frac{(\tau-\xi_j)^2}{2\sigma_j^2}\right] \exp(i\,b_j),
\quad j=1,2
\label{eq:4pAnsatz}
\end{equation}
where height $a$, center position $\xi$, width $\sigma$, and phase $b$ 
are the variational parameters.  
The ansatz was selected as the simplest localized waveform with freedom
to move left or right in order to capture, in the simplest sense,
a possible asymmetry in the solution of the original LL model.
Applying the NCVA method with this, arguably 
over-simplified, four-parameter ansatz, 
leads, through the Euler-Lagrange equations, to a system of algebraic 
differential equations for which the derivatives of the variational
parameters cannot be solved for explicitly. Nonetheless, it is
possible to obtain algebraic equations for the corresponding 
steady state ($\dot{a} = \dot{b} = \dot{\xi} = \dot{\sigma} = 0$)
solutions of the form:
\begin{align}
\begin{cases}
\frac{a^2\sqrt{\pi}}{2 \sigma^2}+\frac{1}{4} a^4\sqrt{2 \pi}+\Delta a^2\sqrt{\pi} &= \frac{a\sin(b)T_0^2 \beta}
{(T_0^2+2\sigma^2)}, \\[1em]
2a^2\sigma\sqrt{\pi} &=  a \cos(b) \sigma \beta, \\[1em]
-\frac{a\sqrt{\pi}}{\sigma}+a^3\sqrt{2\pi}\sigma+2\Delta a\sigma\sqrt{\pi} &= \sin(b) \sigma \beta, \\[1em]
0 &= \frac{-a\xi\sin(b)\beta}{\sigma},
\end{cases}
\label{4pE}
\end{align}
where $\beta = {2 T_0\sqrt{2 \pi X}}/{\sqrt{T_0^2 + 2 \sigma^2}}$.  

Figure~\ref{figSSB4p} depicts the comparison of the bifurcation diagrams
for steady state solutions obtained from the original LL 
model~(\ref{LugiatoLefever}) and the NCVA approach for $\Delta =0.92$ 
and $T_0=2.3$ by monitoring $|u (\tau=0)|^2$ as a function of pump 
peak power $X$, in line with the earlier work of Ref.~\cite{XuCoen}.
Both solutions for the original LL model and the algebraic NCVA system
are obtained by numerical continuation using a standard fixed point iteration 
(Newton-Krylov).
The solutions for the LL model are depicted by the thick curves
while the corresponding NCVA approximations by the thin curves.
Solid and dashed correspond, respectively, to stable and unstable 
solutions.
The insets in the figure depict pulse temporal intensity profiles 
$|u|^2$ for $X = 4$ (symmetric), $X=8$ (symmetric and asymmetric), 
and $X=11$ (symmetric) for both the LL model (thick curves)
and the NCVA reconstructions (thin curves).
For completeness, the insets also show the corresponding linearization
spectra.
As it is evident from the figure, the NCVA with a four-parameter 
ansatz agrees very well with the symmetric branch of LL model
(see black and blue curves). However, for the asymmetric branch
there seems to be a large discrepancy between the LL model and
its NCVA approximation. In fact, 
the 
asymmetric NCVA branch is unstable while it is stable for the original 
LL model. Upon further inspection (details are omitted for brevity), 
the instability of the asymmetric NCVA branch stems, instead of
a pitchfork bifurcation, from a Hopf bifurcation that creates a stable 
limit cycle in the variational parameters.
%

\begin{figure}[htb!]
\centering
\includegraphics[width=4.2cm]{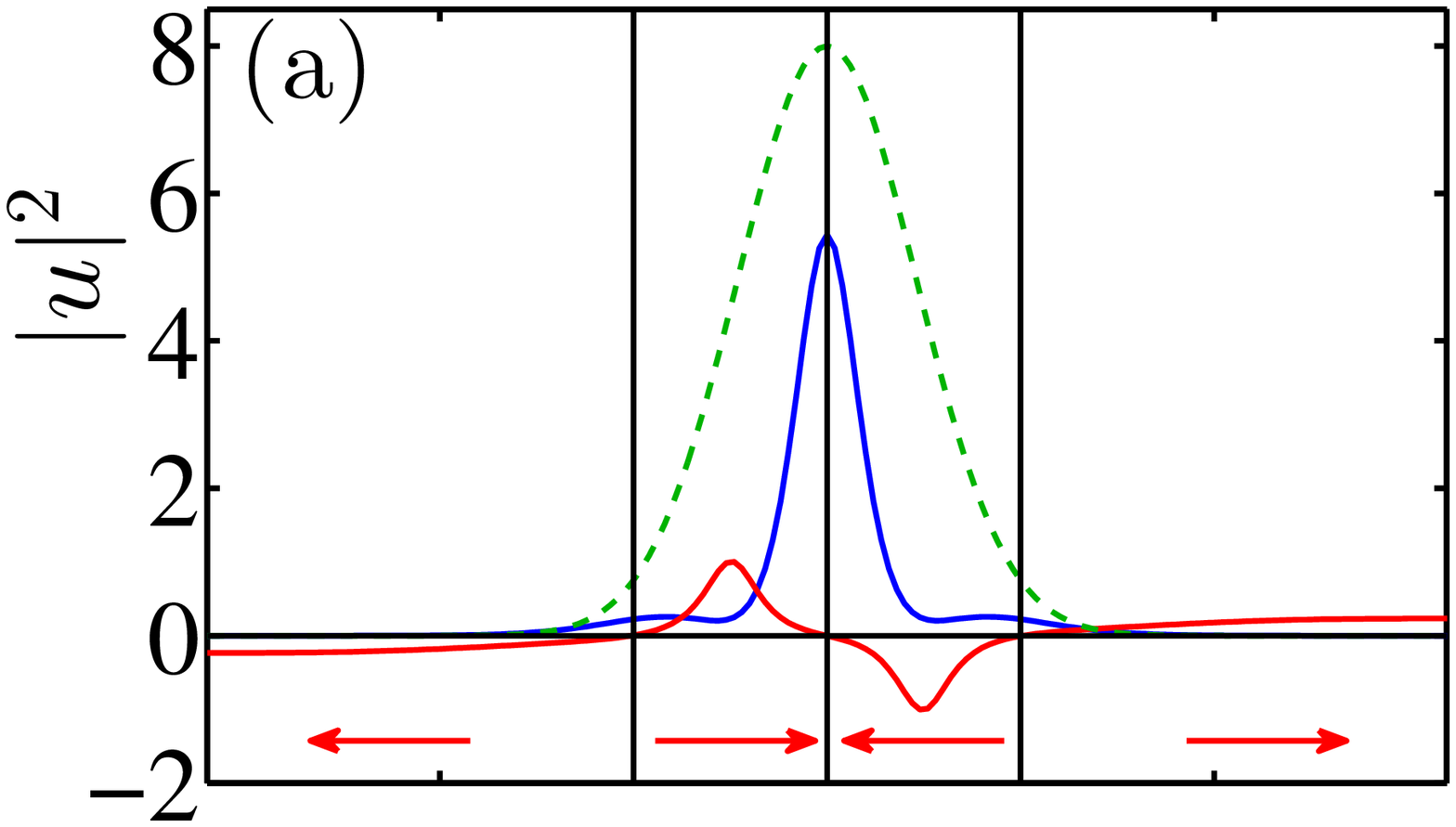}
\includegraphics[width=4.2cm]{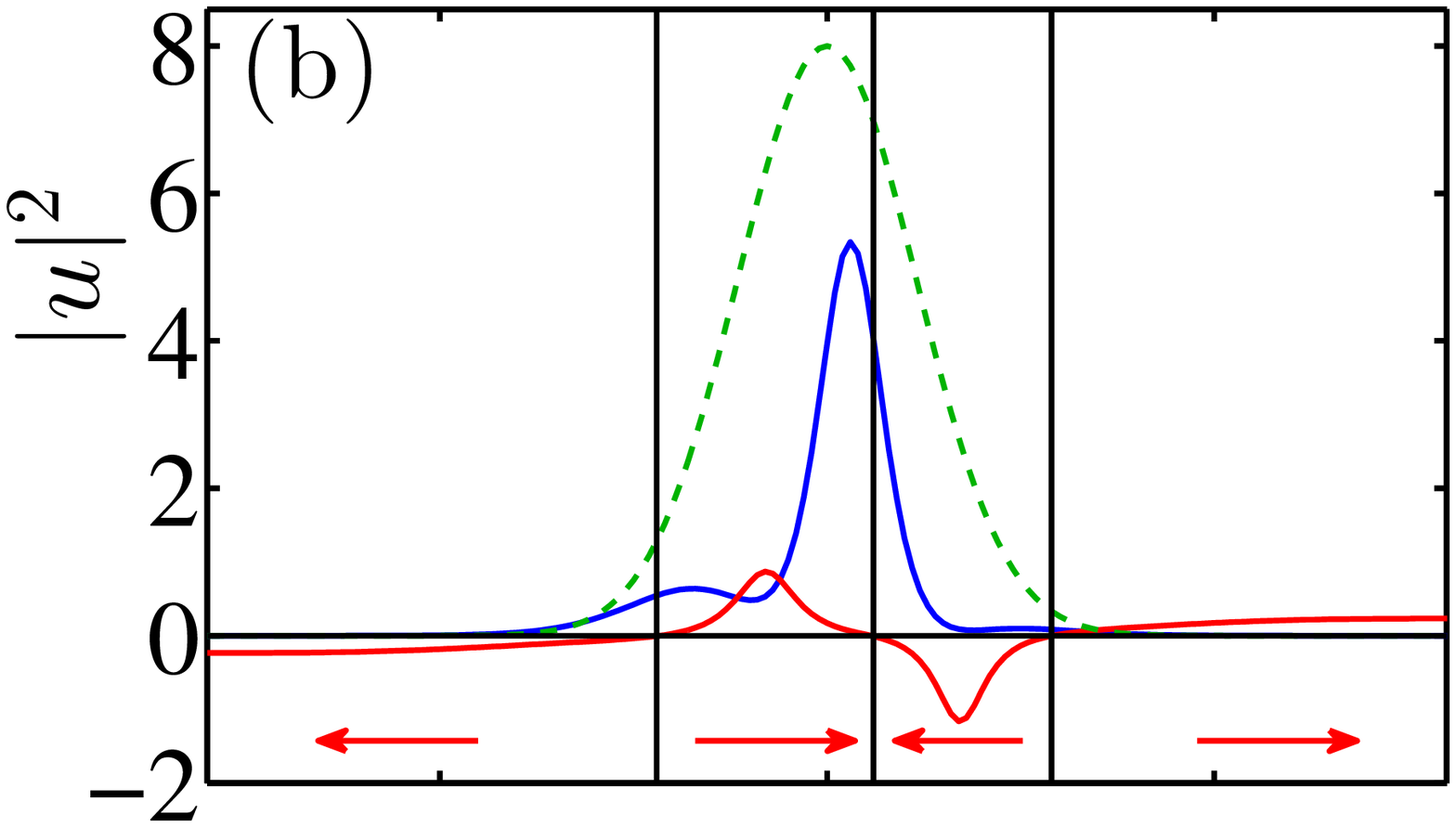}
\\
\includegraphics[width=4.2cm]{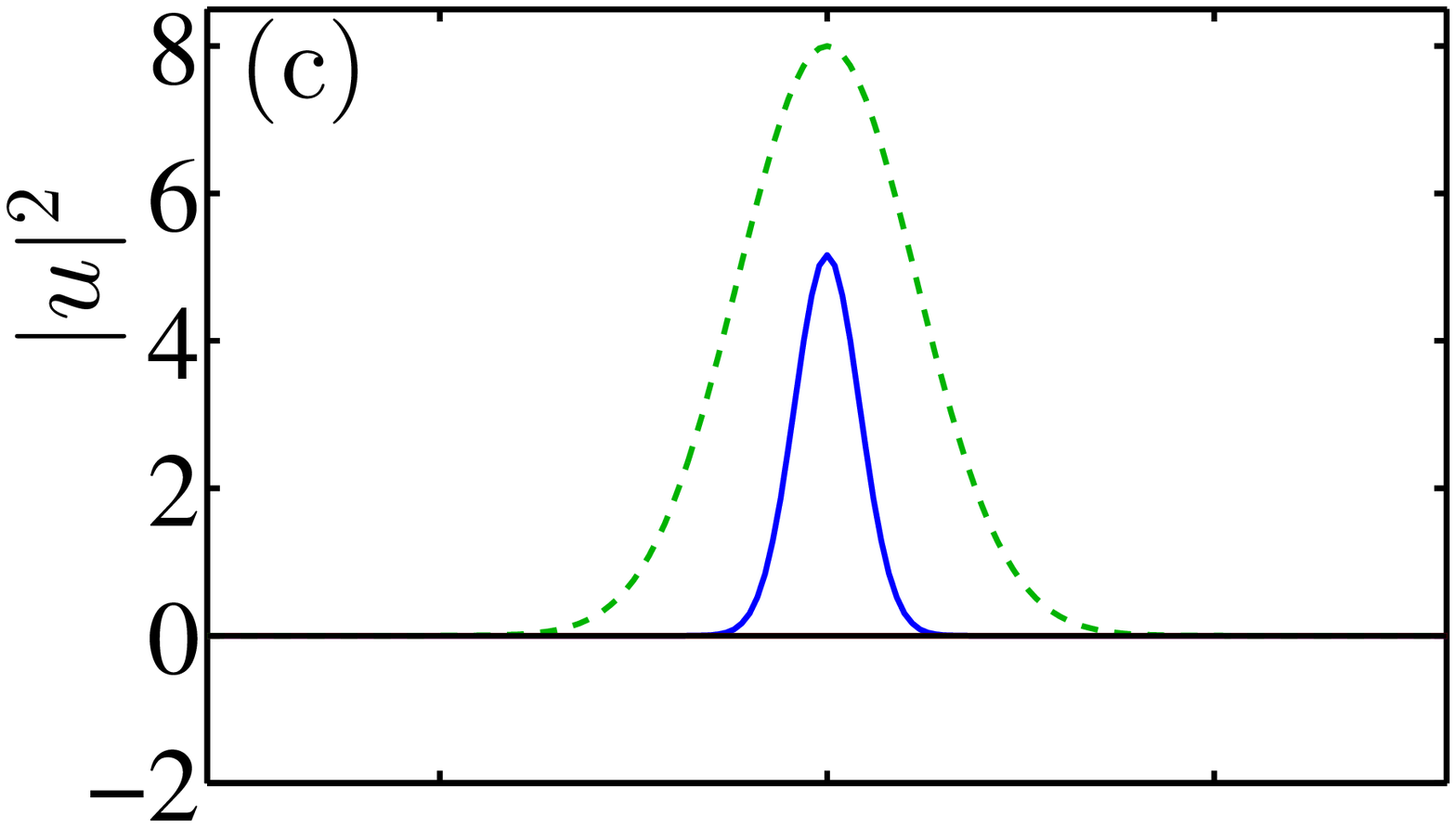}
\includegraphics[width=4.2cm]{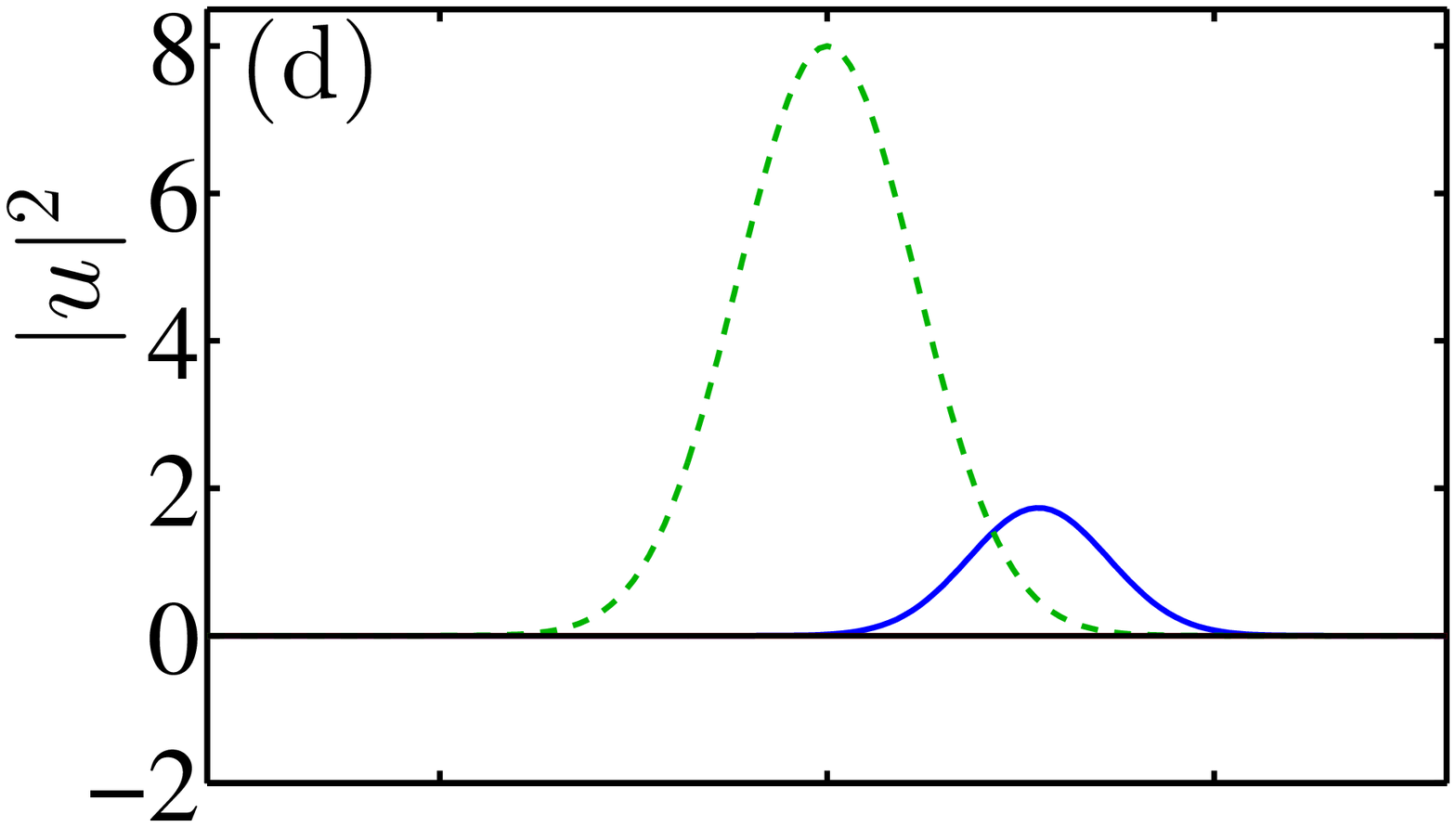}
\\
\includegraphics[width=4.2cm]{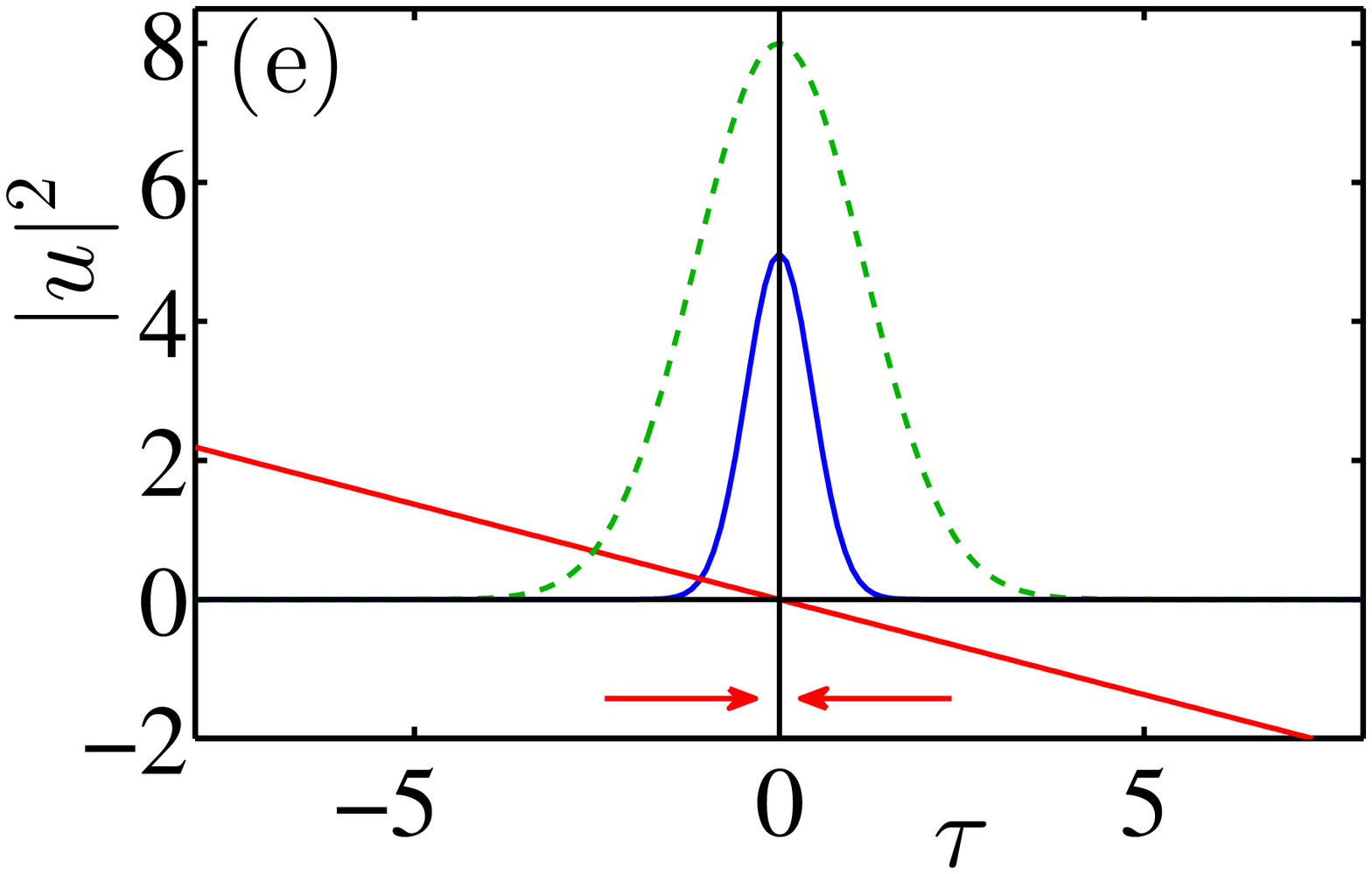}
\includegraphics[width=4.2cm]{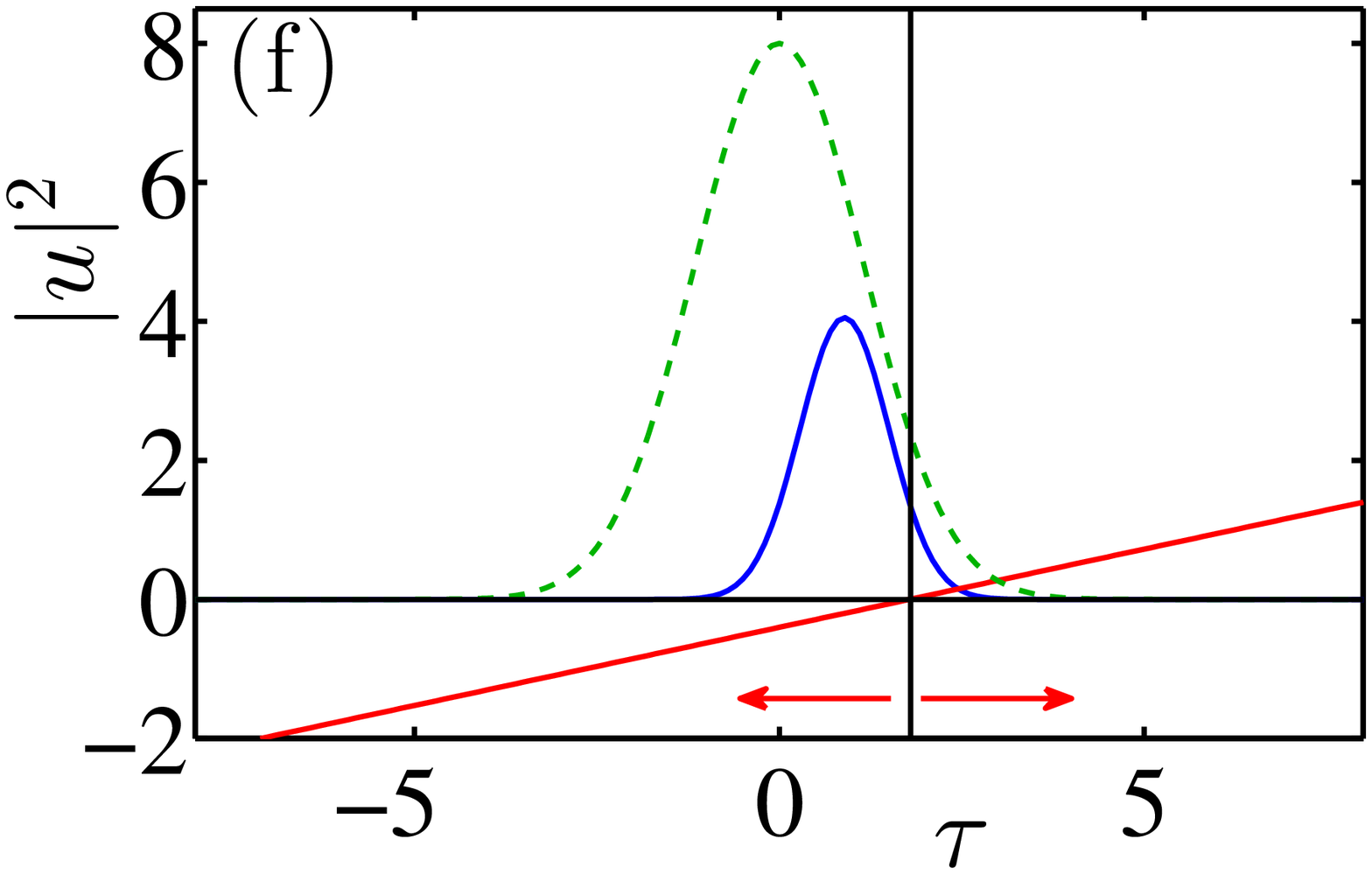}
\caption[]{Understanding the importance of including fluid velocity
terms in the NCVA ansatz. The different panels depict an overlay of 
(i) the temporal intensity profile of the pump pulse $|S(\tau)|^2$ 
(green dashed line),
(ii) the pulse intensity profile $|u|^2$ (blue), and 
(iii) its corresponding fluid velocity (red, $\times 50$)
for 
(a,b) the LL model (top), 
(c,d) the four-parameter NCVA (middle), and 
(e,f) the six-parameter NCVA (bottom)
for symmetric $X=8$ (left) and asymmetric $X=8$ (right) steady state solutions.
The (red) arrows indicate the direction of the fluid velocity which drives 
the SSB bifurcation between the symmetric and asymmetric solutions.
}
\label{fig4SSB}
\end{figure}

It is evident that the four-parameter ansatz is unable to predict the
existence of the pitchfork loop of the original LL model and, furthermore,
although it is able to predict a SSB bifurcation, it fails to give an 
accurate estimation for its threshold (i.e., the critical
pump power needed to observe asymmetric states).
However, this over-simplified ansatz gives two valuable insights regarding 
how to make a more judicious choice for our ansatz. 
Firstly, the ansatz~(\ref{eq:4pAnsatz}) has an inherent complication in that
its corresponding Euler-Lagrange equations lead to a degenerate system
of differential-algebraic equations which can only be explicitly written
for the steady state. 
This degeneracy can be circumvented, as we will show below, by proper balancing 
of the variational parameters in an ansatz with more degrees of freedom.
Secondly, and more importantly, the four-parameter ansatz, by construction, 
only corresponds to real solutions (up to a global phase shift) that lack a 
$\tau$-dependence on their phase.
This lack of $\tau$-dependence on the phase is responsible for the
ansatz solution's lack of internal flow of the field $u$
along the $\tau$-direction.%
\footnote{We remind the reader that when transforming the
NLS equation through the Madelung transformation
$u=\sqrt{\rho}\,e^{i\phi}$ (i.e., writing the wavefunction in
terms of its density $\rho$ and phase $\phi$), one obtains
an evolution equation for the density that corresponds to
an inviscid Eulerian fluid (incorporating the so-called
quantum pressure term that is not important for the current
argument) with the fluid velocity $v$ given precisely by $v=\nabla \phi$.
Thus, the fluid velocity for the system can be obtained by computing
the gradient of the phase of the solution at hand.
Therefore, a lack of a phase variation in the solution
implies a lack of internal flow of the solutions.}
As we explain below, the asymmetric solution is supported
by a delicate balance of the internal flow within this
steady state solution. 
The presence of the underlying flow is clear after careful examination
of the (numerically) exact solutions of the original LL model as depicted
in panels (a) and (b) of Fig.~\ref{fig4SSB}. These panels depict the
density (blue) and phase (red) of the solution where the arrows
indicate the regions where the fluid velocity, as defined by the
gradient of the phase, has different directions. 
The central density maximum, for both the symmetric and
asymmetric solutions, stems from an inward flow towards the
center [which corresponds to a sink of flow due to high density 
through the loss term $iu$ in Eq.~(\ref{LugiatoLefever})]
while the ``wings'', again for both symmetric and asymmetric solutions, 
are supported by sources of the underlying flow maintained by
the pump.
In contrast, the NCVA four-parameter ansatz (\ref{eq:4pAnsatz}) lacks
a phase profile and, therefore, lacks any internal flow as depicted
in panels (c) and (d) of Fig.~\ref{fig4SSB}.
It is then clear that the four-parameter ansatz (\ref{eq:4pAnsatz})
should be inadequate for capturing the important effects of
the underlying current flows of the solutions.

\begin{figure}[htb!]
\centering
\includegraphics[width=8.5cm]{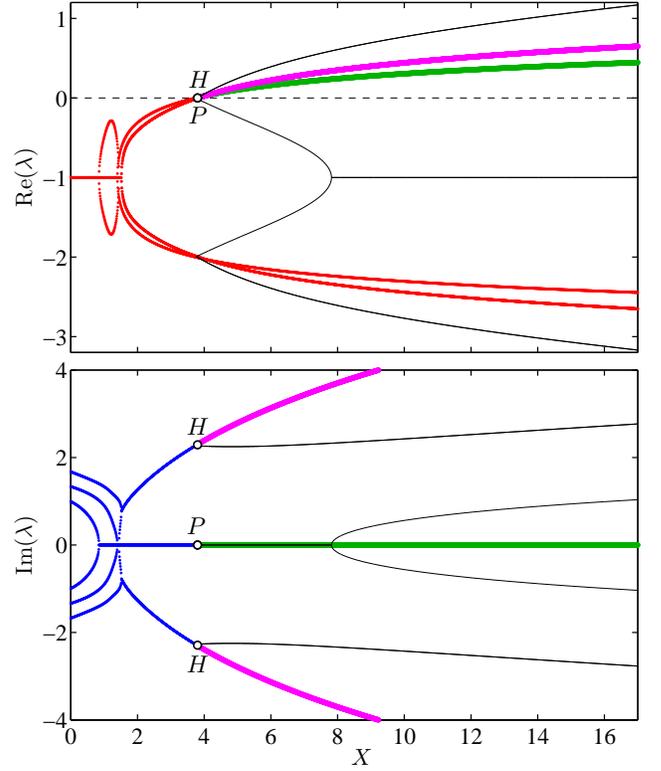}
\caption{
Linearization spectrum for the reduced NCVA ODE (\ref{6pE}).
The notation is the same as the spectrum for the original LL model
depicted in Fig.~\ref{fig:frequencySpectrum}.
The reduced ODE model displays a degenerate bifurcation
consisting of simultaneous pitchfork (P) and a Hopf (H)
bifurcations and thus the asymmetric steady state (see thin black
solid lines) is unstable from its inception.
\label{fig:frequencySpectrum_NCVA}
}
\end{figure}

It is important to mention at this stage that steady state solutions
(for the density) in NLS-type settings incorporating loss and gain
terms must necessarily involve underlying flows that carry
``mass'' from the gain regions towards the lossy regions.
Therefore, in these settings, variational methods should be based
on ans\"atze that incorporate the appropriate underlying current.
Inspired by the appreciation of the presence of such underlying flows 
(and their delicate balance in the steady state solutions)
let us choose an ansatz that is capable of supporting such flows.
Based on this observation, we introduce a six-parameters ansatz
of the form:
\begin{eqnarray}
\bar{u}_j &=&  a_j \exp\left[\frac{-(\tau-\xi_j)^2}{2\sigma_j^2}\right] \times \\ \nonumber
&&\exp\left[i(d_j(\tau-\xi_j)^2 + c_j(\tau - \xi_j) + b_j)\right] ,
\label{6pansatz}
\end{eqnarray}
where, in addition to the parameters height $a$, center position $\xi$, 
width $\sigma$, phase $b$, we have also introduced  velocity $c$ and 
chirp $d$ as variational parameters.
Following again the NCVA methodology for this improved ansatz, we 
obtain the following system of ODEs of the variational parameters:
\begin{align}
\begin{cases}
\dot{a} =& \frac{1}{4}\frac{-4a^2\sigma^3\sqrt{\pi}+3\sigma^2I_b-8a^2d\sigma^3\sqrt{\pi}-2I_d}{a\sigma^3\sqrt{\pi}}, \\[1em]
\dot{b} =& -\frac{1}{8}\frac{8a^2\sqrt{\pi}-5a^4\sqrt{2}\sqrt{\pi}\sigma^2-4I_\sigma \sigma^2+6I_a\sigma a-8a^2c^2\sqrt{\pi}\sigma^2}{ a^2\sigma^2\sqrt{\pi}} \\[1em]
	& + \frac{\sigma c I_c-a^2 \Delta\sqrt{\pi}\sigma^2}{a^2\sigma^2\sqrt{\pi}}, \\[1em]
\dot{c} =& -\frac{c I_b+I_{\xi}}{a^2\sigma\sqrt{\pi}}, \\[1em]
\dot{d} =& \frac{1}{4}\frac{4a^2\sqrt{\pi}-16a^2 d^2 \sigma^4\sqrt{\pi}-a^4\sqrt{2}\sqrt{\pi}\sigma^2-4I_s\sigma^2+2I_a\sigma a}
{a^2\sigma^4\sqrt{\pi}}, \\[1em]
\dot{\sigma} =& -\frac{1}{2}\frac{\sigma^2I_b-8a^2 d \sigma^3\sqrt{\pi}-2I_d}{a^2\sigma^2\sqrt{\pi}}, \\[1em]
\dot{\xi} =& \frac{2a^2\sigma c \sqrt{\pi}+I_c}{a^2\sigma\sqrt{\pi}}, 
\end{cases}
\label{6pE}
\end{align}
where the over-dot denotes derivative with respect to $z$.
Although it is possible to explicitly solve for the derivatives
of the variational parameters, the resulting NCVA ODEs are cumbersome
in that they include the terms $I_a, I_b, I_c, I_d, I_{\xi},$ and 
$I_{\sigma}$ which involve integrals that cannot be explicitly evaluated:
\begin{align}
\begin{cases}
I_a =&\displaystyle \int_{-\infty}^{\infty} E \sin(\Phi) \,d\tau, \\[2ex] \nonumber
I_b =& \displaystyle\int_{-\infty}^{\infty} a E \cos(\Phi) \,d\tau, \\[2ex]
\nonumber
I_c =& \displaystyle\int_{-\infty}^{\infty}  a E (\tau-\xi) \cos(\Phi) \,d\tau, \\[2ex]
\nonumber
I_d  = &\displaystyle\int_{-\infty}^{\infty} a E (\tau-\xi)^2 \cos(\Phi) \,d\tau ,\\[2ex] \nonumber
I_\sigma =&\displaystyle\int_{-\infty}^{\infty} \frac{a E}{\sigma^3} \sin(\Phi) \,d\tau,  \\[2ex] \nonumber
I_\xi =&\displaystyle\int_{-\infty}^{\infty} \Big[ aE \left(-2d(\tau-\xi)-c\right) \cos(\Phi)    \\[1em] \nonumber
&\displaystyle \quad \quad+ \frac{a E (\tau-\xi) }{\sigma^2} \sin(\Phi) \Big] \,d\tau, 
\end{cases}
\end{align}
where $\Phi =d(\tau-\xi)^2+c(\tau-\xi)+b$ and $E = 2\sqrt{X} e^{-\frac{ (\tau-\xi)^2}{2 \sigma^2}}e^{-\frac{\tau^2}{T_0^2}} $.
Nonetheless, for our numerical studies it suffices to evaluate numerically
these integrals as we seek stationary states or as we follow the
dynamics of the parameters as $z$ changes.

\begin{figure*}[htb!]
\centering
\includegraphics[width=15cm]{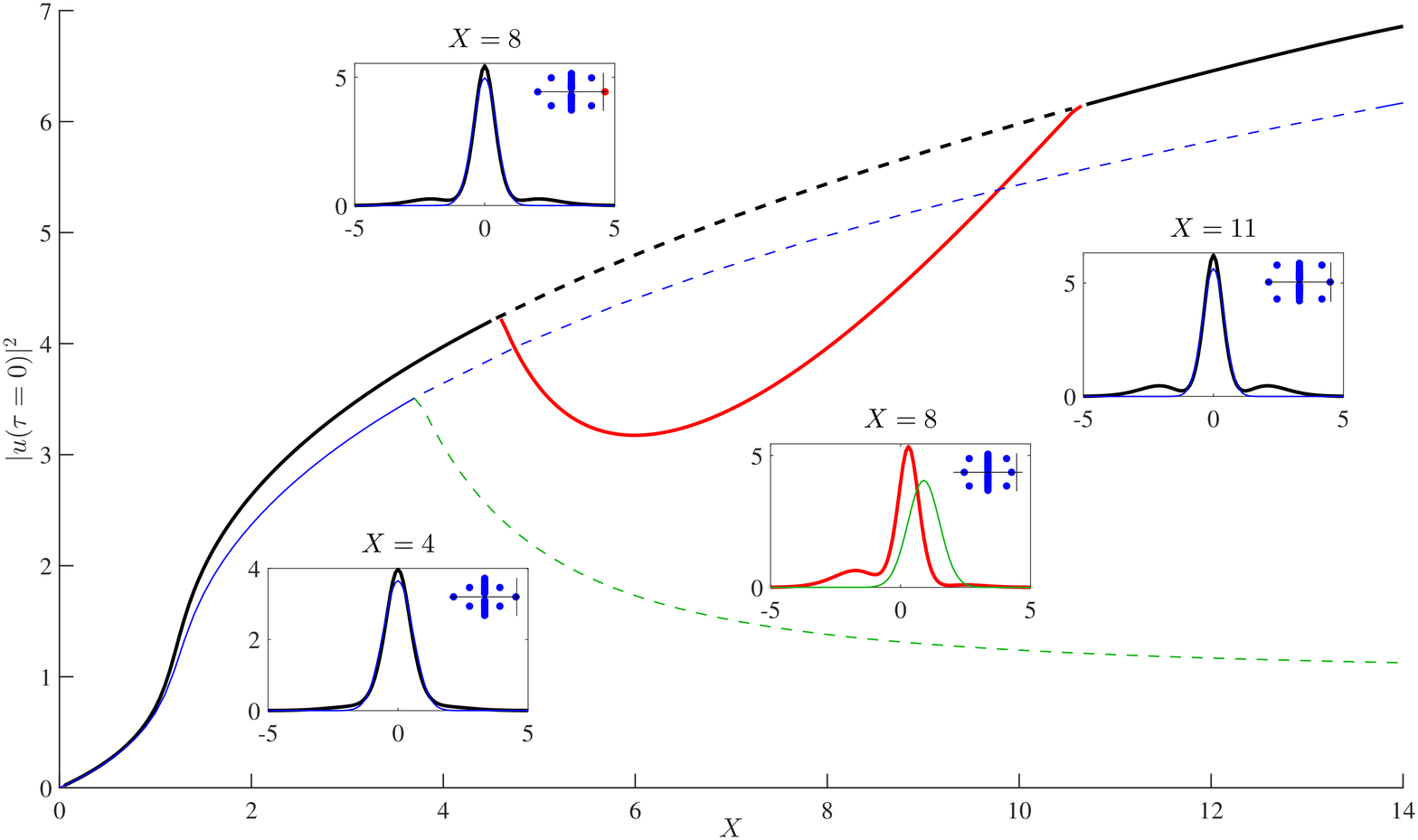}
\caption{Bifurcation diagram as in Fig.~\ref{figSSB4p} but for the more 
complete six-parameter ansatz (\ref{6pansatz}). Same layout and
meaning as in Fig.~\ref{figSSB4p}.}
\label{figSSB6p}
\end{figure*}

Figure \ref{fig:frequencySpectrum_NCVA} depicts the linearization
spectrum for the reduced NCVA six-parameter ODE model (\ref{6pE}) that 
should be compared to the linearization spectrum of original LL model
depicted in Fig.~\ref{fig:frequencySpectrum}.
It is clear that both spectra share some of the bifurcation structure 
but, at the same time, they have notable differences.
For instance, although the reduced NCVA ODE is able to capture
the SSB at $X\approx 3.8$, reasonably close to the actual bifurcation 
of the original model at $X\approx 4.6$, the bifurcation, instead of
being a pitchfork one, is a degenerate one comprising simultaneous 
pitchfork and Hopf bifurcations.
This is the reason why the reduced ODE NCVA model transitions directly
from a stable symmetric solution to a stable (asymmetric) 
limit cycle instead of a stable asymmetric steady state as in 
the original LL model.
The bifurcation diagram for the six-parameter ansatz (\ref{6pansatz}),
together with the one from the original LL model, is depicted in 
Fig.~\ref{figSSB6p} using the same layout as Fig.~\ref{figSSB4p}.  
As it is clear for comparing Figs.~\ref{figSSB4p} and \ref{figSSB6p},
the six-parameter ansatz does a much better job at capturing the
asymmetry states (see insets) and the threshold for the primary
SSB bifurcation than its four-parameter counterpart.
However, as we noted above (and similar to the case of the reduced 
four-parameter NCVA ode) the bifurcation predicted by
the NCVA ODEs (\ref{6pE}) produces an unstable asymmetric state
and a stable (asymmetric) limit cycle.
Nonetheless, the six-parameter ansatz is now able to capture the essence 
of the underlying flow as it can be seen from panels (e) and (f)
of Fig.~\ref{fig4SSB}.
Furthermore, and perhaps more importantly, the improved six-parameter 
NCVA approach is also able to predict reasonably well the threshold 
for the pump power for the onset of the SSB bifurcation.
The fact that the NCVA method is insufficient in characterizing the
details of the SSB instability is, arguably, 
the consequence of employing ans\"atze 
that (while remaining tractable) 
lack the proper freedom to include underlying flows that are
akin to the solutions displayed by the original LL model.
For instance, in order to capture the details of the underlying flows
depicted in panels (a) and (b) of Fig.~\ref{fig4SSB} it should be
necessary to include a fluid flow that has, at least, three
zeros and that would entail, if using polynomials as a basis
for expanding the phase, a quartic polynomial (i.e., five
parameters) for the phase.
Such an ansatz would require five phase variational parameters
and five shape (density) parameters leading to a cumbersome system
of ten couple ODEs. Such a venture falls outside of the scope
of the present manuscript.
%

\section{Local Bifurcation Analysis
\label{secGalerkin}}

In this section, we employ a complementary, dynamical systems inspired
approach based on a 
center manifold reduction to determine 
the dynamics of the system close to the pitchfork bifurcations.

\subsection{Reduced equation}

Let us consider Eq.~(\ref{NLSEq}) with, as before, $\eta=-1$.
In this approach it is more convenient to work with real variables. We therefore set 
\begin{align}
  E = U + iV,
  \label{realvar}
\end{align}
where $U$ and $V$ are real-valued functions, and then Eq.~(\ref{NLSEq})
is equivalent to the system
\begin{align}
\begin{cases}
U_z = -U -V_{\tau\tau} + \Delta V - (U^2 + V^2)V + S(\tau),\\[1.0ex]
V_z = -V + U_{\tau\tau} - \Delta U + (U^2 + V^2) U.
\end{cases}
\end{align}
The numerical computations in Fig.~\ref{figSSB6p} show the existence of a branch of symmetric steady state solutions for values of $X$ between 0 and 14, which can be continued to large $X$. Using the implicit function theorem, one can prove the existence and uniqueness of this branch for small values of $X$, together with the fact that a solution  $(U_X(\tau), V_X(\tau))$ is smooth and decays exponentially to $0$, as $|\tau|\to\infty$. Moreover, one can show that there are no bifurcations, for $X$ sufficiently small.  Here, we are  interested in  the two pitchfork bifurcations predicted by the previous numerical computations at $X_1$ = 4.596695 and $X_2$ = 10.604008.

Consider a symmetric steady solution $(U_X, V_X)$  on the branch in Fig.~\ref{figSSB6p}. By setting
\begin{eqnarray}
\label{pertu}
U = U_X + \nu, \quad \quad V = V_X + v.
\end{eqnarray}
where $\nu$ and $v$ describe the deviations from this steady solution, we 
obtain the new system:
%
\begin{align}
w_z = \mathcal{A}_Xw + \mathcal{J}(\mathcal{R}_{2} (w, w) + \mathcal{R}_3 (w)),
\label{bif1}
\end{align}
in which $w = (\nu, v)^T$ and $\mathcal{A}_X$ is the matrix linear operator 
\begin{align}
\mathcal{A}_X = -\mathbb{I} + \mathcal{J}\mathcal{L}_X,  
\label{Ax}
\end{align}
where $\mathbb{I} $ is the identity matrix,
\begin{align}
\mathcal{J} = \begin{pmatrix} 0 & -1 \\ 1 & 0  \end{pmatrix},
\nonumber
\end{align}
and $\mathcal{L}_X$ is the linear operator defined by
\begin{align}
\mathcal{L}_X = \begin{pmatrix} \partial^2_{\tau} - \Delta + 3 U_X^2 + V_X^2 & 2U_X V_X \\ 2 U_X V_X & \partial^2_{\tau} - \Delta + U_X^2 + 3V_X^2 \end{pmatrix}.
\nonumber
\end{align}
Finally, $\mathcal{R}_{2} (w_1, w_2)$  is the bilinear map given by
\begin{align}
\mathcal{R}_{2} (w_1,w_2) = \begin{pmatrix} 3U_X \nu_1 \nu_2 + V_X (\nu_1 v_2 + \nu_2 v_1) + U_X v_1 v_2 \\
V_X \nu_1 \nu_2 + U_X (\nu_1 v_2 + \nu_2 v_1) + 3 V_X v_1 v_2 \end{pmatrix},
\nonumber
\end{align}
for $w_1 =  (\nu_1, v_1)^T$ and $w_2 = (\nu_2, v_2)^T$, and $\mathcal{R}_{3} (w)$ is the cubic map given by
\begin{align}
\mathcal{R}_3 (w) = \begin{pmatrix} (\nu^2 + v^2)\nu \\ (\nu^2 + v^2)v  \end{pmatrix},
\nonumber
\end{align}
for $w = (\nu, v)^T$.  
We regard Eq.~(\ref{bif1}) as an infinite-dimensional dynamical system in the phase space $H = L^2(\mathbb{R}) \times L^2(\mathbb{R})$ equipped with the usual scalar product
\begin{align}
\langle w_1, w_2 \rangle = \int_{\mathbb R} (\nu_1(\tau)\nu_2(\tau) + v_1(\tau)v_2(\tau)) d\tau.
\nonumber
\end{align}
In this Hilbert space, $\mathcal{A}_X$ is a closed linear operator with domain $H^2(\mathbb{R}) \times H^2 (\mathbb{R})$, and the operators $\mathcal{J}$ and $\mathcal{L}_X$ are skew- and self-adjoint, respectively.  The nonlinear terms $\mathcal{R}_{2}$ and $\mathcal{R}_3$ are smooth maps.

Varying the parameter $X$ in Eq.~(\ref{bif1}), the bifurcation points are the values of $X$ where the structure of the purely imaginary part of the spectrum of the linear operator $\mathcal{A}_X$ changes.
Notice that the spectrum of $\mathcal{A}_X$ is symmetric with respect to the vertical line 
Re$(\lambda) = -1$ in the complex plane. Indeed, since $\mathcal J$ and $\mathcal L_X$ are skew- and self-adjoint operators, respectively, the spectrum of $\mathcal J\mathcal L_X$ is symmetric with respect to the imaginary axis, so that the spectrum of $\mathcal{A}_X =- \mathbb I + \mathcal J\mathcal L_X$ is symmetric with respect to the vertical line 
Re$(\lambda) = -1$ in the complex plane.  Moreover, it is also symmetric with respect to the real axis, since  $\mathcal{A}_X$ is a real operator.
These two properties are clearly satisfied by the numerically obtained
spectrum depicted in Fig.~\ref{fig:frequencySpectrum}.

The essential spectrum of  $\mathcal{A}_X$ can be determined analytically. Since $\mathcal{A}_X$ is a differential operator with asymptotically constant coefficients, its essential spectrum coincides with the spectrum of the asymptotic operator $\mathcal{A}_0$ which has constant coefficients. Then a standard Fourier analysis allows to compute explicitly the spectrum of  $\mathcal{A}_0$, and conclude that the essential spectrum of  $\mathcal{A}_X$ is the set
\begin{align}
\nonumber
\sigma_{\rm ess} = \{ -1 \pm i (k^2 + \Delta),  \; k \in \mathbb{R} \},
\end{align}
which lies entirely in the open left half complex plane. Consequently, bifurcations can only arise due to point spectrum, which consists of eigenvalues with finite algebraic multiplicities. 
For sufficiently small $X$, standard perturbation arguments show that the spectrum of $\mathcal{A}_X$ stays close to the one of $\mathcal{A}_0$.  In particular, it lies in the left half complex plane, and no bifurcations/instabilities occur for small $X$. The previous numerical computations show that there exists a first value $X_1$ at which one (simple) eigenvalue crosses the origin and becomes positive for $X > X_1$
(see point $P_1$ in Fig.~\ref{fig:frequencySpectrum}).  
All other eigenvalues have negative real parts.  Increasing $X$, there is a second value $X_2$ where this simple eigenvalue crosses the origin back in the left half complex plane
(see point $P_2$ in Fig.~\ref{fig:frequencySpectrum}).
Our purpose is to  study the two (pitchfork) bifurcations which occur at 
these parameter values, and which are directly related
to the SSB phenomena observed experimentally in Ref.~\cite{XuCoen}.

We denote by $\lambda_0(X)$ the simple eigenvalue above, so that we have
\begin{align}
\sigma(\mathcal{A}_X) = \{ \lambda_0(X)\} \cup \sigma_- (\mathcal{A}_X), \\
\quad \sigma_-(\mathcal{A}_X) \subset \{\lambda \in \mathbb{C}\, ; \, 
\mathrm{Re} (\lambda) \leqslant -\gamma \},
\end{align}
for some $\gamma > 0$, and
\begin{align}
  \lambda_0(X) < 0 \; \mbox{ for } \; X < X_1,
  \label{lambda0} \\
\lambda_0 (X_1) = 0, \nonumber \\
\lambda_0 (X) > 0 \; \mbox{ for } \; X_1 < X< X_2, \nonumber \\
\lambda_0(X_2) = 0. \nonumber
\end{align}

\begin{figure*}[htb!]
\centering
\includegraphics[width=15cm]{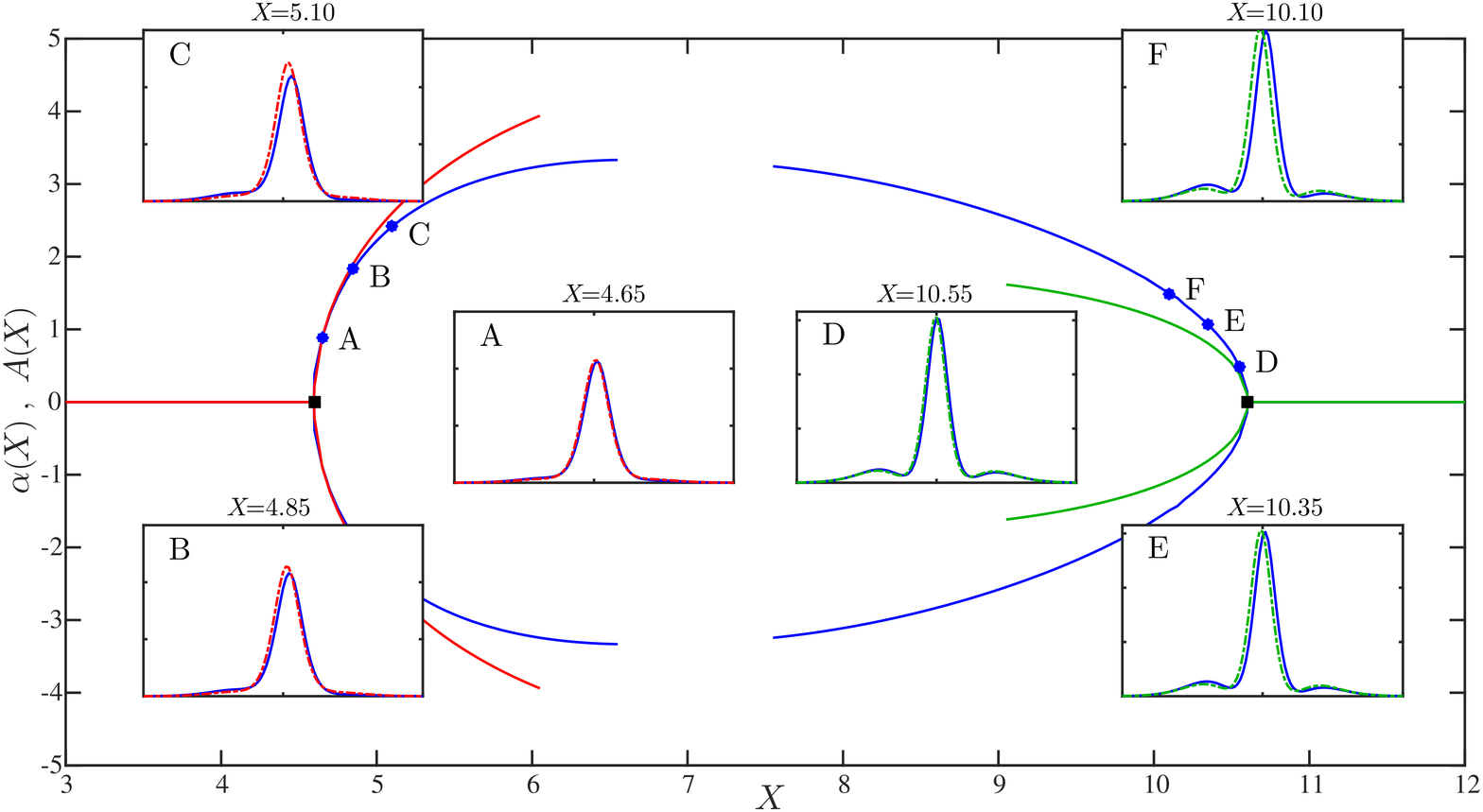}
\caption{
Steady state comparison between the original LL model~(\ref{LugiatoLefever})
and the {center manifold} approach (see text) close to the 
pitchfork bifurcation points.
The figure depicts the coefficients determining the amount of 
asymmetry for the LL model as monitored by $\alpha(X)$,
defined in Eq.~(\ref{PDEfit}), (see blue curves containing
the points A, B, C, D, E, and F) and for the {center manifold} 
approach by $A(X)$
(see red [about bifurcation point $X_1$] and green
[about bifurcation point $X_2$] curves).
The insets correspond to the steady state asymmetric solutions for
both the LL model (solid curves) and the {center manifold} approach
(dashed curves) at the points A, B, C, D, E, and F indicated in
the bifurcating branches corresponding, respectively, to pump powers 
$X$ = 4.65, 4.85, 5.1, 10.55, 10.35, and 10.1.}
\label{PirchforkBifurcationComparison}
\end{figure*}

The following arguments work for both bifurcation points $X_1$ and $X_2$.  Choose one of these values and denote it by $X_*$. We set
\begin{align}
\mathcal{A}_* = \mathcal{A}_{X_*}, \quad 
\mathcal{L}_* = \mathcal{L}_{X_*}, \quad {\rm and} \quad
\mathcal{R}_{*,2} = \mathcal{R}_{X_*, 2}.
\nonumber
\end{align}
Further, we consider an eigenvector $\zeta_*$ in the one-dimensional kernel of  $\mathcal{A}_*$ and an eigenvector $\zeta_*^*$ in the one-dimensional kernel of the adjoint operator $(\mathcal{A}_{*})^*$. We claim that we can choose $\zeta_*^*$ such that
\begin{align}
\zeta_*^* = \mathcal{J} \zeta_2 \; \mbox{~and~}  \;  \mathcal{A}_{*} \zeta_2 = -2 \zeta_2.
\nonumber
\end{align}
Indeed, since  $\mathcal J$ and $\mathcal L_X$ are skew- and self-adjoint operators, respectively, we find
\begin{align}
(\mathcal{A}_{*})^* \zeta_*^* = 0 \, &\Leftrightarrow \, - (\mathcal{L}_*  \mathcal{J}) \zeta_*^* = \zeta_*^*, \\
&\Leftrightarrow \, \mathcal{A}_{*} (\mathcal{J}\zeta_*^* ) = -2 (\mathcal{J} \zeta_*^*).
\nonumber
\end{align}
The last equality shows that $\mathcal{J}\zeta_*^*$ is an eigenvector of $\mathcal{A}_{*}$ associated to the eigenvalue $-2$ [the symmetric of 0 with respect to the vertical line Re$(\lambda) = -1$], and proves the claim.

The analytical and numerical computations of the essential and point spectra, respectively, above show that $\mathcal{A}_*$ has precisely one simple eigenvalue on the imaginary axis [located at the origin], and that the remaining spectrum lies entirely in the open left half complex plane. By arguing with the center manifold theorem, e.g., see \cite[Chapter 2]{MH-Iooss}, we conclude that the dynamical system~(\ref{bif1}) possesses a one-dimensional center manifold, for any $X$ close to $X_*$. All bounded solutions of  Eq.~(\ref{bif1}) lie on this manifold and are of the form
\begin{align}
w(z) = A(z)\zeta_* + \Phi (A(z), X), 
\label{centermanifold}
\end{align}
in which $A$ is a real-valued function
and $\Phi$, depending upon $A$ and the parameter $X$, satisfies
\begin{align}
\Phi(A,X) = \mathcal{O} (|A| (|X - X_*| + |A|),
\nonumber
\end{align}
for small $A$ and $X$ close to $X_*$, and the orthogonality condition
\begin{align}
\langle \Phi(A, X), \zeta_*^* \rangle = 0.
\label{ortho}
\end{align}
Here, $\zeta_*$ and  $\zeta_*^*$ are the eigenvectors in the kernels of the operator $\mathcal{A}_*$ and its adjoint operator, respectively.
The dynamics of the center manifold is determined by a scalar ODE
\begin{align}
\frac{dA}{dz} = f(A, X). 
\label{scalarODE}
\end{align}

Our purpose is to compute the leading order terms in the expansion of the reduced scalar field $f$.
Notice that the system~(\ref{bif1}) is invariant under the reflection $\tau \mapsto - \tau$.  As a consequence, $f$ is odd in $A$,
\begin{align}
f(A,X) = - f(-A,X),
\nonumber
\end{align}
so that its Taylor expansion is of the form
\begin{align}
f(A,X) = c_0 (X) A + c_3 A^3 + \mathcal{O} (|A|^3 (|X-X_*| + A^2) ),
\nonumber
\end{align}
in which $c_0(X)$ and $c_3$ are real constants. Since the (single) eigenvalue of the linearization at $0$ of the reduced scalar field $f$ is precisely $\lambda_0(X)$, we conclude that 
\begin{align}
c_0(X) = \lambda_0 (X),
\nonumber
\end{align}
In particular $c_0(X_*) = 0 $ and its values for $X$ close to $X_*$ are given by the previous numerical calculations (see Sec.~\ref{secModel}).  Next, in order to compute $c_3$, we set $X=X_*$ and replace the ansatz~(\ref{centermanifold}) into the system~(\ref{bif1}). Taking into account Eq.~(\ref{scalarODE}), the expansion of $f$, and expanding the reduction function $\Phi(A,X)$ at $X=X_*$,
\begin{align}
\Phi(A,X_*) = \Phi_0 A + \Phi_2 A^2 + \Phi_3 A^3+\mathcal{O} (|A|^4),
\nonumber
\end{align}
we obtain the equality
\begin{align}
\mathcal{A}_{*} \Phi_2 = - \mathcal{J} R_{*,2}(\zeta_*, \zeta_*).
\end{align}
The orthogonality condition Eq.~(\ref{ortho}) determines uniquely $\Phi_2$, and as a consequence of the reflection symmetry $\tau \mapsto - \tau$ of Eq.~(\ref{bif1}), we have that $\Phi_2$ is an even function. In particular, $\Phi_2$ is the unique even solution of the equation. Next, at $\mathcal{O}(A^3)$, we obtain:
\begin{align}
c_3  \zeta_* = \mathcal A_*\Phi_3 + 2\mathcal{J} \mathcal{R}_{*, 2}(\zeta_*, \Phi_2) +  \mathcal{J} \mathcal{R}_3 (\zeta_*).
\nonumber
\end{align}
Taking the scalar product with $\zeta_*^*= \mathcal{J} \zeta_2$, and using the fact that $\zeta_*^*$ belongs to the kernel of the adjoint of $\mathcal A_*$,  we obtain the second coefficient
\begin{align}
c_3 = \frac{1}{\langle \zeta_*, \mathcal{J} \zeta_2 \rangle} \left( \langle 2 \mathcal{R}_{*,2}(\zeta_*, \Phi_2), \zeta_2 \rangle + \langle \mathcal{R}_3(\zeta_*), \zeta_2 \rangle \right).
\nonumber
\end{align}

\subsection{Local dynamics}

The local dynamics on the one-dimensional center manifold is qualitatively given by the signs of the two coefficients $c_0 (X)$ and $c_3$. These coefficients are computed numerically, and the result confirms the situation depicted in Fig.~\ref{figSSB6p}. The sign of $c_0(X)$ is the same as the one of $\lambda_0(X)$ [see Eq.~\ref{lambda0}], and
\begin{align}
  c_3<0, \; \mbox{~for~} \;  X=X_1, \mbox{~and~}  \;
  c_3>0, \; \mbox{~for~}  \; X=X_2.
\end{align}
At both bifurcation points $X_1$ and $X_2$, we are in the presence of a pitchfork bifurcation in which a pair of asymmetric stable equilibria appears from or disappears into
the symmetric equilibrium which, in turn, changes its stability. Moreover, there is a pair of heteroclinic orbits connecting the unstable symmetric equilibrium at $z=-\infty$ with the stable asymmetric equilibria at $z=\infty$. These solutions persist for the full system, and can be computed as solutions of  Eq.~(\ref{NLSEq}) going back through the reduction procedure, successively from the formulas~(\ref{centermanifold}), (\ref{pertu}), and (\ref{realvar}). In particular, the heteroclinic connection describes the transition dynamics from the unstable symmetric to the stable asymmetric solution.

\begin{figure}[htb!]
\centering
\includegraphics[width=8.5cm]{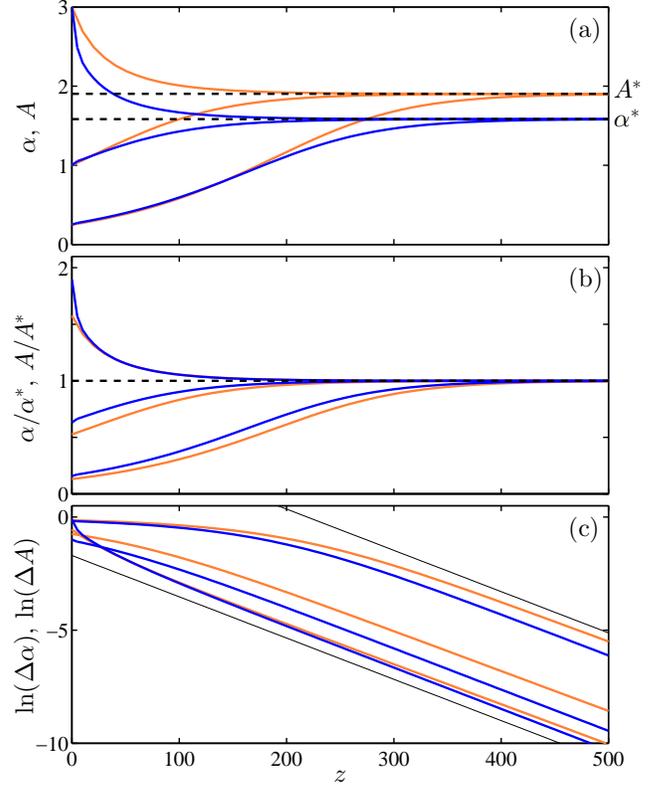}
\caption{Orbits representing the dynamics settling to the
asymmetric steady states past the first pitchfork bifurcation point.
Depicted is the evolution for the asymmetry coefficients 
$\alpha(z)$ and $A(z)$ (see text) for a 
pump strength is $X=4.85$ that is to the right of the
first pitchfork bifurcation point $X_1=4.596695$.
The blue (dark) curves correspond to the original model
($\alpha(z)$)  while the orange (gray) curves correspond to the 
{reduced equation} ($A(z)$).
The orbits tend towards their corresponding steady state solutions
$\alpha^*$ and $A^*$ which correspond to the stable asymmetric
state created by the pitchfork bifurcation.
Panel (b) corresponds to panel (a) by normalizing the $A(z)$
and $\alpha(z)$ orbits by their respective steady states.
Panel (c) shows the logarithm of the normalized distance to the
steady state $\Delta \alpha=(\alpha-\alpha^*)/\alpha^*$ 
and $\Delta A=(A-A^*)/A^*$.
{%
In this panel we also depict with thin (black) lines the slope 
$\lambda(4.85)=-0.01824$ corresponding 
to the stability eigenvalue of the asymmetric state [see
small black dot next to the point $P_1$ on the thin (black) 
branch depicted in the top panel of Fig.~\ref{fig:frequencySpectrum}]
which is shown to coincide with the rate of attraction towards
the asymmetric steady {state} for both the original model and {the reduced equation}.
}
}
\label{PitchPhasePortrait}
\end{figure}

Based on the bifurcation analysis above, {we compare the solutions~(\ref{centermanifold}) given by the center manifold 
approach to ones found directly from} the original LL model~(\ref{NLSEq}) 
for $\eta = -1$, $\Delta = 0.92$, and $T_0 = 2.3$.  
In particular, we compare the asymmetric stationary states described by 
Eqs.~(\ref{pertu}) and~(\ref{centermanifold}) with the ones obtained from 
the LL~(\ref{LugiatoLefever})
by projecting the numerically found steady state solutions of
the latter along the symmetric and asymmetric branches for values 
of $X$ near the bifurcation points $X_1$ and $X_2$.  
Therefore, the asymmetric solutions of Eq.~(\ref{LugiatoLefever}) are 
fit using the symmetric solutions plus $\alpha$ times the
eigenvector of the translation mode, i.e. the eigenvector $\zeta_*$ 
associated with the bifurcation point $X_*$ [see Eq.~(\ref{PDEEigenProblem}) 
and Fig.~\ref{fig:frequencySpectrum}], but now written in 
original complex variables associated with $u(z,\tau)$.

Therefore, we find the best (in the least-squares sense) scalar
value $\alpha$ such that:
\begin{align}
u_{\mathrm{Asym} }(X_* + \delta X)\approx u_{\mathrm{Sym} }(X_* + \delta X) + 
\alpha(X_* + \delta X)\, \zeta_*.
\label{PDEfit}
\end{align}
By using a nonlinear least-square solver, we extract the value of 
$\alpha(X)$ around each bifurcation point $X_1$ and $X_2$ and compare it
with the value of $A(X)$ from {the reduced equation}.
Fig.~\ref{PirchforkBifurcationComparison} depicts a plot of
$A(X)$ and $\alpha(X)$ close to both pitchfork bifurcations.  
As the figure shows, the shape of the bifurcation is 
well captured by the {center manifold} approach. In fact, as expected, 
the {reduced equation} correctly captures the concavity of the 
bifurcating branch at both bifurcation points.
In the figure, the insets depict the steady state profile comparison 
between the original model and the {center manifold} approach for values
of the pump power $\delta X$ = 0.05, 0.25, and 0.5 units away
from both bifurcation points.
As it is clear from the insets, the {center manifold} approach approximates
very well the shape of the steady state solutions particularly close
to the bifurcation points.
Therefore, the {reduced equation provides} a very good agreement 
for the {\em statics}, i.e.~steady states, of the original model
close to the bifurcation points.

We now focus on the dynamics close to the bifurcation.
In particular, let us study how solutions, starting from
a perturbed (unstable) symmetric solution, evolve towards the
(stable) asymmetric steady [an example portraying this evolution
is depicted in Fig.~\ref{fig:evolution}(a)].
Figure~\ref{PitchPhasePortrait}(a) depicts the dynamical evolution
of the asymmetry coefficients $A$ and $\alpha$, as defined above,
for initial conditions above and below the corresponding
steady state solutions $A^*$ and $\alpha^*$ for a value
of $X$ past the first pitchfork bifurcation point.
As the figure shows, both the original LL dynamics and the
{center manifold reduction} produce orbits that settle towards
their corresponding (stable) asymmetric steady states.
In order to better compare the decay in both systems,
we depict in Fig.~\ref{PitchPhasePortrait}(b) the orbits
normalized by their corresponding steady states.
Finally in Fig.~\ref{PitchPhasePortrait}(c) we depict the
logarithm of the distance to the corresponding steady states.
As it is clear from this panel, the steady state is reached
exponentially fast with a rate that precisely coincides with
the stability eigenvalue for the asymmetric state
{%
[see small black dot next to the point $P_1$ on the thin (black) branch 
depicted in the top panel of Fig.~\ref{fig:frequencySpectrum}]
as suggested by the thin black (dark) lines depicting the
rate using $\lambda(4.85)=-0.01824$.}
The figure confirms that the {center manifold} approach is not only
capable of reproducing the right statics for the asymmetric
branches, but it is also capable of reproducing the main qualitative
features of the dynamics
as the solutions settle towards the stable asymmetric states.

\section{Conclusions \& Future Challenges
\label{secConclusion}}
In this paper we considered different theoretical techniques
aiming at a more detailed analytical and numerical understanding
of the phenomenology arising in a 
coherently-driven passive optical Kerr resonator, experimentally
observed in Ref.~\cite{XuCoen} and modelled by the 
Lugiato-Lefever equation (LL)~\cite{LL} that corresponds to a 
non-Hamiltonian variant of the nonlinear Schr\"odinger equation.
In particular, we  applied both a non-conservative variational approximation
(NCVA) of Ref.~\cite{JuliaNCVA} and a {center manifold} technique to
study the spontaneous symmetry breaking (SSB) bifurcations arising
in this system.
It is found that variational ans\"atze lacking the appropriate 
phase variation are not able to capture the intrinsic underlying
velocity fields and the delicate
balance present in the steady state density solution.
These flows are ubiquitous in systems with gain and loss as
the steady state consists of a balance between regions with
gain and loss provided by flows from the former regions (sources) 
to the latter ones (sinks).
Using a suitably adjusted variational ansatz, including higher order
phase terms while remaining tractable, 
the NCVA is capable of accurately predicting the
threshold in the pump power for the onset of SSB
---although it is not adequate for fully capturing the complex
bifurcation structure (especially so at large pumping strength/large
nonlinearity). 
To obtain a more complete and quantitative, as well as 
mathematically a more rigorously justifiable description, 
we have then employed a center manifold approach 
capable of capturing both the
forward and reverse pitchfork bifurcations of the
original system in terms of the corresponding locations and 
profile shapes of the steady states
and also in terms of the rate of convergence towards the
stable asymmetric state when the symmetric one is rendered
unstable. The numerical determination of the linearization spectrum
of the system was not only important for completing the calculations
associated with the {center manifold} method; it was also crucial towards
a detailed understanding of the full stability/instability
transitions.

In that same vein, 
the identification of the parametric dependence of the spectrum has
enabled us to uncover the emergence in the original LL model of
 a (potentially quite relevant to experiments in this system) 
Hopf bifurcation. This, in turn, 
was dynamically found to give rise to stable periodic solutions
and hence illustrate that more complex bifurcation scenaria may arise
as the cavity loss parameter is varied.

It should be interesting to study in more detail these
more complex bifurcation and SSB scenaria and their implications 
for the original physical system. In that regard, it may be beneficial
to explore the possibility to identify these periodic orbits as
exact solutions of the numerical LL problem past the Hopf bifurcation
point that we have identified here, via a fixed point iteration
at the Poincar{\'e} recurrence of the relevant periodic orbit,
{or using the center manifold technique}.
Moreover, this would enable to explore the stability (Floquet
multipliers) associated with this orbit. Another natural direction
would be to consider similar LL models in two-dimensional
settings (even if these may be less relevant from an 
experimental perspective in nonlinear optics) in order to
appreciate how SSB phenomena may interplay with external drives
and also with the potential of such higher dimensional models
to feature collapse. Lastly, from the point of view of more
recent experiments in connection to the LL equation, a deeper
understanding of the dynamics and interactions, as well as the
trapping and manipulation of temporal cavity solitons 
(and corresponding effective ``particle'' descriptions thereof)
may be relevant to pursue~\cite{tweezing,patterns}.

\section*{Acknowledgements}
J.R.~gratefully acknowledges the support from the Computational Science
Research Center (CSRC) at SDSU, the ARCS foundation and Cymer.
R.C.G.~gratefully acknowledges the support of NSF-DMS-1309035.
P.G.K.~gratefully acknowledges the support of
NSF-DMS-1312856 and from
the ERC under FP7, Marie Curie Actions, People,
International Research Staff Exchange Scheme (IRSES-605096).
M.H.~gratefully acknowledges the support of
the LabEx ACTION (project AMELL) and the
ANR project BoND (ANR-13-BS01-0009-01).

\section*{References}

\end{document}